\documentstyle[12pt,epsfig,rotate,rotating]{article}

\newcommand{\tka}{\mbox{$\tau^- \!\!\rightarrow\!\! \,\nu_\tau\, {\rm K}_1(1270)$}}
\newcommand{\tkb}{\mbox{$\tau^- \!\!\rightarrow\!\! \,\nu_\tau\, {\rm K}_1(1400)$}}

\newcommand{\tppx}{\mbox{$\tau^- \!\!\rightarrow\!\! \,\nu_\tau\, \pi^-\pi^-{\rm X}^+$}}

\newcommand{\tkpp}{\mbox{$\tau^- \!\!\rightarrow\!\! \,\nu_\tau\, {\rm K}^-\pi^-\pi^+$}}
\newcommand{\tkpk}{\mbox{$\tau^- \!\!\rightarrow\!\! \,\nu_\tau\, {\rm K}^-\pi^-{\rm K}^+$}}
\newcommand{\tppk}{\mbox{$\tau^- \!\!\rightarrow\!\! \,\nu_\tau\, \pi^-\pi^-{\rm K}^+$}}
\newcommand{\tkkp}{\mbox{$\tau^- \!\!\rightarrow\!\! \,\nu_\tau\, {\rm K}^-{\rm K}^-\pi^+$}}

\newcommand{\tppxzz}{\mbox{$\tau^- \!\!\rightarrow\!\! \,\nu_\tau\, \pi^-\pi^-{\rm X}^+ (\pi^0)$}}

\newcommand{\tkppz}{\mbox{$\tau^- \!\!\rightarrow\!\! \,\nu_\tau\, {\rm K}^-\pi^-\pi^+ \pi^0$}}
\newcommand{\tkpkz}{\mbox{$\tau^- \!\!\rightarrow\!\! \,\nu_\tau\, {\rm K}^-\pi^-{\rm K}^+ \pi^0$}}

\newcommand{\tpppzz}{\mbox{$\tau^- \!\!\rightarrow\!\! \,\nu_\tau\, \pi^-\pi^-\pi^+ (\pi^0)$}}
\newcommand{\tkppzz}{\mbox{$\tau^- \!\!\rightarrow\!\! \,\nu_\tau\, {\rm K}^-\pi^-\pi^+ (\pi^0)$}}
\newcommand{\tkpkzz}{\mbox{$\tau^- \!\!\rightarrow\!\! \,\nu_\tau\, {\rm K}^-\pi^-{\rm K}^+ (\pi^0)$}}
\newcommand{\tppkzz}{\mbox{$\tau^- \!\!\rightarrow\!\! \,\nu_\tau\, \pi^-\pi^-{\rm K}^+ (\pi^0)$}}

\newcommand{\tkkzz}{\mbox{$\tau^- \!\!\rightarrow\!\! \,\nu_\tau\, {\rm K}^-{\rm K}^0_{\rm S} (\pi^0)$}}

\newcommand{\tmu}{\mbox{$\tau^- \!\!\rightarrow\!\! \mu^- {\overline{\nu}_\mu} \,\nu_\tau$}} 
 
\newcommand{\eemm}{\mbox{${\rm e}^+{\rm e}^- \!\!\rightarrow\! {\rm e}^+{\rm e}^-\mu^+\mu^-$}} 
\newcommand{\epem}{\mbox{${\rm e}^+{\rm e}^-$}} 
\newcommand{\dimuon}{\mbox{${\rm e}^+{\rm e}^- \!\!\rightarrow\! \mu^+\mu^-$}} 
\newcommand{\eett}{\mbox{${\rm e}^+{\rm e}^- \!\!\rightarrow\! \tau^+\tau^-$}} 
\newcommand{\kpp}{\mbox{$\nu_\tau\, {\rm K}^-\pi^-\pi^+$}}
\newcommand{\kpk}{\mbox{$\nu_\tau\, {\rm K}^-\pi^-{\rm K}^+$}}

\newcommand{\kppb}{\mbox{${\rm K}^-\pi^-\pi^+$}}
\newcommand{\kpkb}{\mbox{${\rm K}^-\pi^-{\rm K}^+$}}

\newcommand{\kppz}{\mbox{$\nu_\tau\, {\rm K}^-\pi^-\pi^+\pi^0$}}
\newcommand{\kpkz}{\mbox{$\nu_\tau\, {\rm K}^-\pi^-{\rm K}^+\pi^0$}}

\newcommand{\kppzz}{\mbox{$\nu_\tau\, {\rm K}^-\pi^-\pi^+(\pi^0)$}}
\newcommand{\kpkzz}{\mbox{$\nu_\tau\, {\rm K}^-\pi^-{\rm K}^+(\pi^0)$}}
\newcommand{\ppkzz}{\mbox{$\nu_\tau\, \pi^-\pi^-{\rm K}^+(\pi^0)$}}
\newcommand{\kkpzz}{\mbox{$\nu_\tau\, {\rm K}^-{\rm K}^-\pi^+(\pi^0)$}}
\newcommand{\kkkzz}{\mbox{$\nu_\tau\, {\rm K}^-{\rm K}^-{\rm K}^+(\pi^0)$}}

\newcommand{\de}{\mbox{${\rm d}E/{\rm d}x$}}
\newcommand{\dmeas}{\mbox{$D_{\rm meas}$}}
\newcommand{\dpred}{\mbox{$D_{\rm pred}$}}

\newcommand{\sbeta}{\mbox{$s(\beta)$}}   
\newcommand{\sbetap}{\mbox{$s^\prime(\beta)$}}   
\newcommand{\sres}{\mbox{$s_{res}$}}
\newcommand{\fphi}{\mbox{$f(\phi)$}}

\newcommand{\ekpp}{\mbox{$\epsilon_{{\rm K}^-\pi^-\pi^+}$}}
\newcommand{\ekpk}{\mbox{$\epsilon_{{\rm K}^-\pi^-{\rm K}^+}$}}
\newcommand{\eppk}{\mbox{$\epsilon_{\pi^-\pi^-{\rm K}^+}$}}
\newcommand{\ekppz}{\mbox{$\epsilon_{{\rm K}^-\pi^-\pi^+\pi^0}$}}
\newcommand{\ekpkz}{\mbox{$\epsilon_{{\rm K}^-\pi^-{\rm K}^+\pi^0}$}}
\newcommand{\eppkz}{\mbox{$\epsilon_{\pi^-\pi^-{\rm K}^+\pi^0}$}}
\newcommand{\bkpp}{\mbox{${\rm B}_{{\rm K}^-\pi^-\pi^+}$}}
\newcommand{\bkpk}{\mbox{${\rm B}_{{\rm K}^-\pi^-{\rm K}^+}$}}

\newcommand{\bkppz}{\mbox{${\rm B}_{{\rm K}^-\pi^-\pi^+\pi^0}$}}
\newcommand{\bkpkz}{\mbox{${\rm B}_{{\rm K}^-\pi^-{\rm K}^+\pi^0}$}}

\newcommand{\bkppzz}{\mbox{${\rm B}_{{\rm K}^-\pi^-\pi^+(\pi^0)}$}}
\newcommand{\bkpkzz}{\mbox{${\rm B}_{{\rm K}^-\pi^-{\rm K}^+(\pi^0)}$}}

\newcommand{\kpke}{${K^-\pi^-K^+}$-enhanced}
\newcommand{\kpks}{${K^-\pi^-K^+}$-depleted}

\newcommand{\etal}{{\it et al.}}
\newcommand{\mkpp}{\mbox{$M_{{\rm K}^-\pi^-\pi^+}$}}
\newcommand{\mkp}{\mbox{$M_{{\rm K}^-\pi^+}$}}
\newcommand{\mpp}{\mbox{$M_{\pi^-\pi^+}$}}
\newcommand{\mkpk}{\mbox{$M_{{\rm K}^-\pi^-{\rm K}^+}$}}
\newcommand{\mkk}{\mbox{$M_{{\rm K}^-{\rm K}^+}$}}
\newcommand{\mpk}{\mbox{$M_{\pi^-{\rm K}^+}$}}
\newcommand{\mh}{\mbox{hadronic ${\rm Z}^0$}}
\newcommand{\faz}{\mbox{$f_{{\rm K}_1(1400)}$}}
\newcommand{\fbz}{\mbox{$f_{{\rm K}_1(1270)}$}}
\newcommand{\fcz}{\mbox{$f_{{\rm K}^\ast \pi}$}}
\newcommand{\fdz}{\mbox{$f_{\rho {\rm K}}$}}

\begin{document}
\sloppy
\begin{titlepage}
\begin{center}{\large   EUROPEAN LABORATORY FOR PARTICLE PHYSICS
}\end{center}\bigskip
\begin{flushright}
CERN-EP/99-095 \\ July 8, 1999
\end{flushright}

\vspace{0.0cm}

\begin{center}
 {\LARGE\bf A Study of Three-Prong\\
          Tau Decays with Charged Kaons \\
       \  \\
 }
\vspace{0.5cm}
  {\Large\bf The OPAL Collaboration} \\
\end{center}
 
\begin{abstract}
From an analysis of the ionisation energy loss
of charged particles selected from
a sample of 147926 \eett\ candidates recorded in the OPAL
detector at \epem\ centre-of-mass energies near the ${\rm Z}^0$
resonance, 
we determine the branching ratios:
\begin{eqnarray*}
{\rm Br}(\tau^-\rightarrow \,\nu_\tau\, {\rm K}^-\pi^-\pi^+ (\pi^0)) 
     & = & 0.343\pm0.073\pm0.031 \;\% 
\nonumber \\[1mm] 
 {\rm Br}(\tau^-\rightarrow \,\nu_\tau\, {\rm K}^-\pi^-{\rm K}^+ (\pi^0))  
     & = & 0.159\pm0.053\pm0.020 \;\% ,
\nonumber 
\end{eqnarray*}
where the $(\pi^0)$ notation refers to decay modes with or without
an accompanying $\pi^0$.
The \tkppzz\ final states occurring through \tkkzz\
are treated as background in this analysis.

We also examine the resonant structure of \tkpp\ candidates.
Under the assumption that the resonant
structure is dominated by the K$_1$ resonances,
we determine:
\begin{eqnarray*}
 R & = & {{{\rm Br}(\tau^-\rightarrow \,\nu_\tau\, {\rm K}_1(1270))}\over
 {
{\rm Br}(\tau^-\rightarrow \,\nu_\tau\, {\rm K}_1(1400))
+
{\rm Br}(\tau^-\rightarrow \,\nu_\tau\, {\rm K}_1(1270))}
} 
      =   0.71 \pm 0.16 \pm 0.11 .
\nonumber
\end{eqnarray*}
In all results,
the first uncertainties are statistical and the second are systematic.
 
\end{abstract}


\begin{center}{\large
(Submitted to European Physical Journal C)
}\end{center}
\end{titlepage}

\begin{center}{\Large        The OPAL Collaboration
}\end{center}\bigskip
\begin{center}{
G.\thinspace Abbiendi$^{  2}$,
K.\thinspace Ackerstaff$^{  8}$,
G.\thinspace Alexander$^{ 23}$,
J.\thinspace Allison$^{ 16}$,
K.J.\thinspace Anderson$^{  9}$,
S.\thinspace Anderson$^{ 12}$,
S.\thinspace Arcelli$^{ 17}$,
S.\thinspace Asai$^{ 24}$,
S.F.\thinspace Ashby$^{  1}$,
D.\thinspace Axen$^{ 29}$
% ...
}\end{center}\bigskip
\bigskip
$^{  1}$School of Physics and Astronomy, University of Birmingham,
Birmingham B15 2TT, UK
\newline
$^{  2}$Dipartimento di Fisica dell' Universit\`a di Bologna and INFN,
I-40126 Bologna, Italy
\newline
% ...
$^{  8}$CERN, European Organisation for Particle Physics,
CH-1211 Geneva 23, Switzerland
\newline
$^{  9}$Enrico Fermi Institute and Department of Physics,
University of Chicago, Chicago IL 60637, USA
\newline
% ...
$^{ 12}$Indiana University, Department of Physics,
Swain Hall West 117, Bloomington IN 47405, USA
\newline
% ...
$^{ 16}$Department of Physics, Schuster Laboratory, The University,
Manchester M13 9PL, UK
\newline
$^{ 17}$Department of Physics, University of Maryland,
College Park, MD 20742, USA
\newline
% ...
$^{ 23}$Department of Physics and Astronomy, Tel Aviv University,
Tel Aviv 69978, Israel
\newline
$^{ 24}$International Centre for Elementary Particle Physics and
Department of Physics, University of Tokyo, Tokyo 113-0033, and
Kobe University, Kobe 657-8501, Japan
\newline
% ...
$^{ 29}$University of British Columbia, Department of Physics,
Vancouver BC V6T 1Z1, Canada
\newline
% ...
 Debrecen, Hungary.
\newline

\section{Introduction}
\leavevmode\indent
The study of rare decays of the tau lepton has been
made possible by the availability of large, low background
tau-pair samples, such as those produced at the Z$^0$ resonance
at LEP I.  This paper reports on an analysis of three-prong
tau decay modes containing charged kaons, using the complete set of data
collected by the OPAL experiment between 1990 and 1995
at \epem\ centre-of-mass energies near the ${\rm Z}^0$ 
resonance.
The excellent charged particle identification capability of the OPAL
detector is exploited to study these states and obtain
measurements of their branching ratios. 
 
The $\tau^\pm$ lepton can only decay to hadronic final states
of strangeness 0 or $\pm1$.
These two allowed values of the final state strangeness 
imply 
that 
if a first-order weak decay of a tau to a three-prong final
state includes exactly one kaon, then
that kaon, if it is charged, must have the same
charge as the initial tau.  Likewise, if a first-order weak decay of a tau to 
a three-prong final state includes exactly two charged kaons, 
those kaons must be of opposite charge.  
Thus \tkpp\ and \tkpk\ decays can both
occur as first-order weak interactions.      
Decays which violate the strangeness
constraint, such as 
\tppk\ and \tkkp, 
can only occur as second
order weak interactions, and are therefore
highly suppressed.
The charge conjugate decays are implied in these interactions and
throughout this paper.

Table \ref{tab:tab1a} lists three-prong decays
of the tau with charged kaons in the final state, 
along with the current experimental measurements and theoretical
expectations for each decay branching ratio
\cite{bib:aleph}-\cite{bib:theory3}.
Due to the relative rarity of tau three-prong final
states with charged kaons, and the experimental difficulties
involved in charged particle identification, the branching ratios
and resonant
structure of these decays are not well known.
We determine the branching ratios for
\tkppzz\ and \tkpkzz, 
and, to verify that
the charged particle identification techniques
used in this analysis yield unbiased branching ratio
measurements, we also determine the branching ratio
to the \tppkzz\ final state to ensure that the result is
consistent with zero.
In addition, we examine the resonant
structure of these decays, and, under the assumption that
the \tkpp\ final state is dominated by the K$_1$ intermediate
resonances, we determine the branching ratio of \tka\ relative
to the sum of the \tka\ and \tkb\ branching ratios.
 
The \tkppzz\ final states occurring through \tkkzz\ 
are considered background in the analysis presented here.  
In addition,
three-prong tau decays with charged kaons and
more than one $\pi^0$ are severely phase-space
suppressed and are therefore neglected.        
 
\section{The OPAL Detector}
\leavevmode\indent
The OPAL detector is described in detail in reference
\cite{bib:opaldet}.  The component of the detector most
important to this analysis is the central
jet chamber,
which measures the momentum and
specific ionisation energy loss, \de, 
of charged particles. The jet chamber provides the only
means of distinguishing between charged pions and kaons in the
momentum range of interest to this study.
 
The jet chamber is a cylinder
4 m long and 3.7 m in diameter, and is divided 
by cathode wire planes        
into 24 azimuthal sectors.  
Each sector contains one radial plane of anode wires, which are
staggered to resolve left-right ambiguities.
The chamber is
contained in a solenoidal 
magnetic field of $0.435$ T,
and is filled with an argon-methane-isobutane gas mixture at a pressure
of 4 atmospheres.  This arrangement provides a transverse spatial
resolution of $\sigma_{xy} = 130$ $\mu$m, and a two-hit resolution
of $2.5$ mm.
In the barrel region of the
jet chamber ($|\cos{\theta}| < 0.72$) \footnote{A spherical coordinate
system is used, with the $+z$-axis in the direction of the circulating
electron beam.  The angle $\theta$ is defined as the
polar angle with respect to the $+z$-axis, and $\phi$ is defined
as the azimuthal angle measured from the $+x$-axis, which points
towards the centre of the LEP ring.},\hspace{1mm} the ionisation energy
loss of a charged
particle is sampled up to 159 times.
A truncated mean is formed by discarding
the highest $30\%$ of the measurements, 
resulting in a \de\
resolution of about $3\%$ for isolated tracks in the 
chamber~\cite{bib:dedxref}.
 
A layer of wire chambers with drift direction parallel to the
$z$-axis lies immediately outside the jet chamber.  
These
$z$-chambers
accurately determine the polar angle of charged particles
traversing the central detector.
A lead-glass electromagnetic calorimeter and presampler
chambers are located outside the magnetic coil and jet chamber
pressure vessel.  
The electromagnetic calorimeter is primarily used in
this analysis to identify tau decays which include a
$\pi^0$ in the final state.
The return yoke of the OPAL magnet is instrumented for
hadron calorimetery, and is surrounded by external muon chambers.
The hadron calorimeter and muon chambers are mainly
used in this analysis to select control samples containing
muons and to veto non-tau backgrounds in the
tau-pair candidate sample.
 
\subsection{Parameterisation of Ionisation Energy Loss}
\begin{figure}[p]
   \begin{center}
     \mbox{ \epsfxsize=15cm
            \epsffile[23 165 524 652]{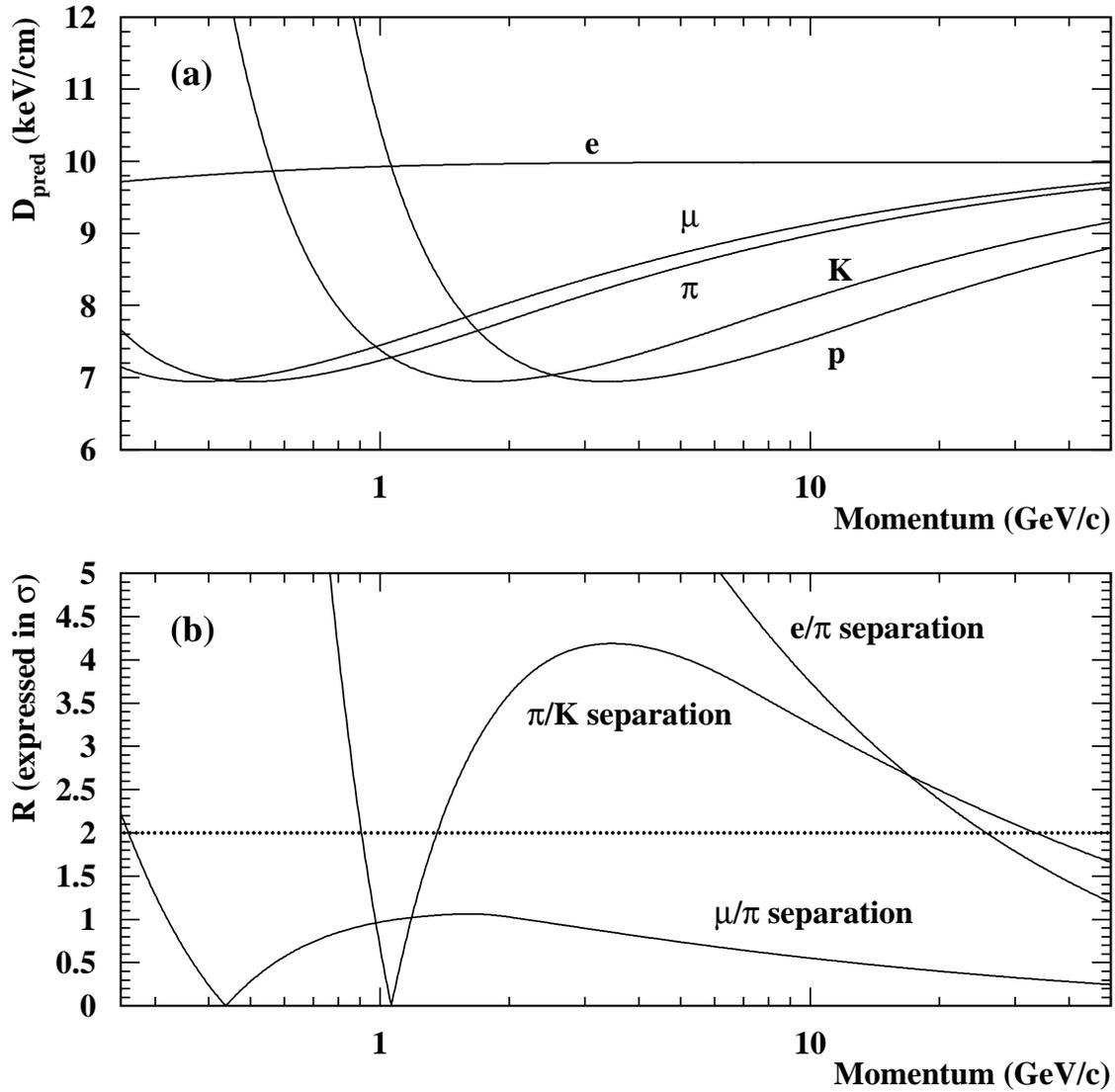} }
   \end{center}
 \vspace*{0.5cm}
 \caption[foo]{
 \label{fig:predict}
 (a) shows the ionisation energy loss \dpred\ as a function
 of the momentum for various particles in the OPAL jet chamber.
 (b) shows the resolution power ${\cal R}_{ij}$ 
 expressed in
 terms of the \de\ resolution $\sigma$,
 for various pairs of particle species $i$ and $j$.
 }
 \end{figure}

\leavevmode\indent
 Figure \ref{fig:predict}(a) shows the dependence of \de\
 on the momentum of various particles in the OPAL jet chamber,
 as predicted by the
 OPAL energy loss parameterisation, \dpred~\cite{bib:dedxref}.
 Figure \ref{fig:predict}(b) shows the
 particle resolving power, ${\cal{R}}_{ij}$,
 versus momentum for various pairs of particle species $i$ and $j$, 
 where 
\begin{eqnarray}
{\cal{R}}_{ij} \equiv |D^i_{\rm pred} - D^j_{\rm pred}| \, / \, \sigma  ,
\nonumber
\end{eqnarray}
 and where $\sigma$ is the mean of the uncertainties of $D^i_{\rm pred}$
 and $D^j_{\rm pred}$.
 Note that the OPAL jet chamber
 yields a pion/kaon separation of at least
 $2\sigma$ for particles between about 2 to 30 GeV/c.

The parameters used in the calculation of \dpred\ and
$\sigma$ 
are tuned to the measured energy loss of charged particles in \mh\
decays, yielding a \de\ parameterisation that is accurate enough
for nearly all analyses of OPAL data.
 
However, hadronic tau decays are characterised by lower multiplicity and
narrower jets than those produced in \mh\ decays, 
resulting in small systematic differences
between the
measured \de\ in the two environments.  Although these differences
are less than one $\sigma$ in magnitude, they must be corrected
for accurate particle identification in three-prong tau decays.  
Based on a study of the measured energy loss of 
muons in \dimuon,\tmu, and \eemm\ samples,
the following corrections are applied to the
measured energy loss, \dmeas, and to the OPAL 
\de\ parameterisation and its uncertainty, \dpred\ and $\sigma$:
\begin{description}
\item[{\rm Multiplicative correction to \dpred\ :}]
      To obtain a parameterisation which 
      correctly predicts the measured 
      energy loss of particles from tau decays, \dpred\
      must be corrected with a multiplicative factor, $s(\beta)$.
      Parameterising $s(\beta)$ as a second
      order polynomial in ${Q(\beta)= -\log(1-\beta^2)}$ yields 
      satisfactory results.  
      The magnitude of the
      correction is of order $1\%$ for tracks in one-prong tau decays.
\item[{\rm Additive correction to \dmeas\ :}]
      The measured energy loss of charged
      particles in low-multiplicity events is found
      to depend strongly on 
      the azimuthal separation, $\phi$, between the track and
      anode plane in the jet chamber cell.
      To improve the \de\ resolution, this behaviour is
      corrected with a function, $f(\phi)$, with
      seven parameters tuned to the
      measured energy loss of \dimuon\ candidates.
      The magnitude of this correction is also of order $1\%$.
\item[{\rm Multiplicative correction to $\sigma$:}]
      To correctly
      predict the spread in the \de\ of charged particles
      from one-prong tau decays,
      the OPAL parameterisation of $\sigma$ must be
      corrected
      with a multiplicative factor, $s_{\rm res}\approx 0.9$.  
\end{description}
 
\begin{figure}[ht]
   \begin{center}
     \mbox{ \epsfxsize=15cm
            \epsffile[37 161 524 649]{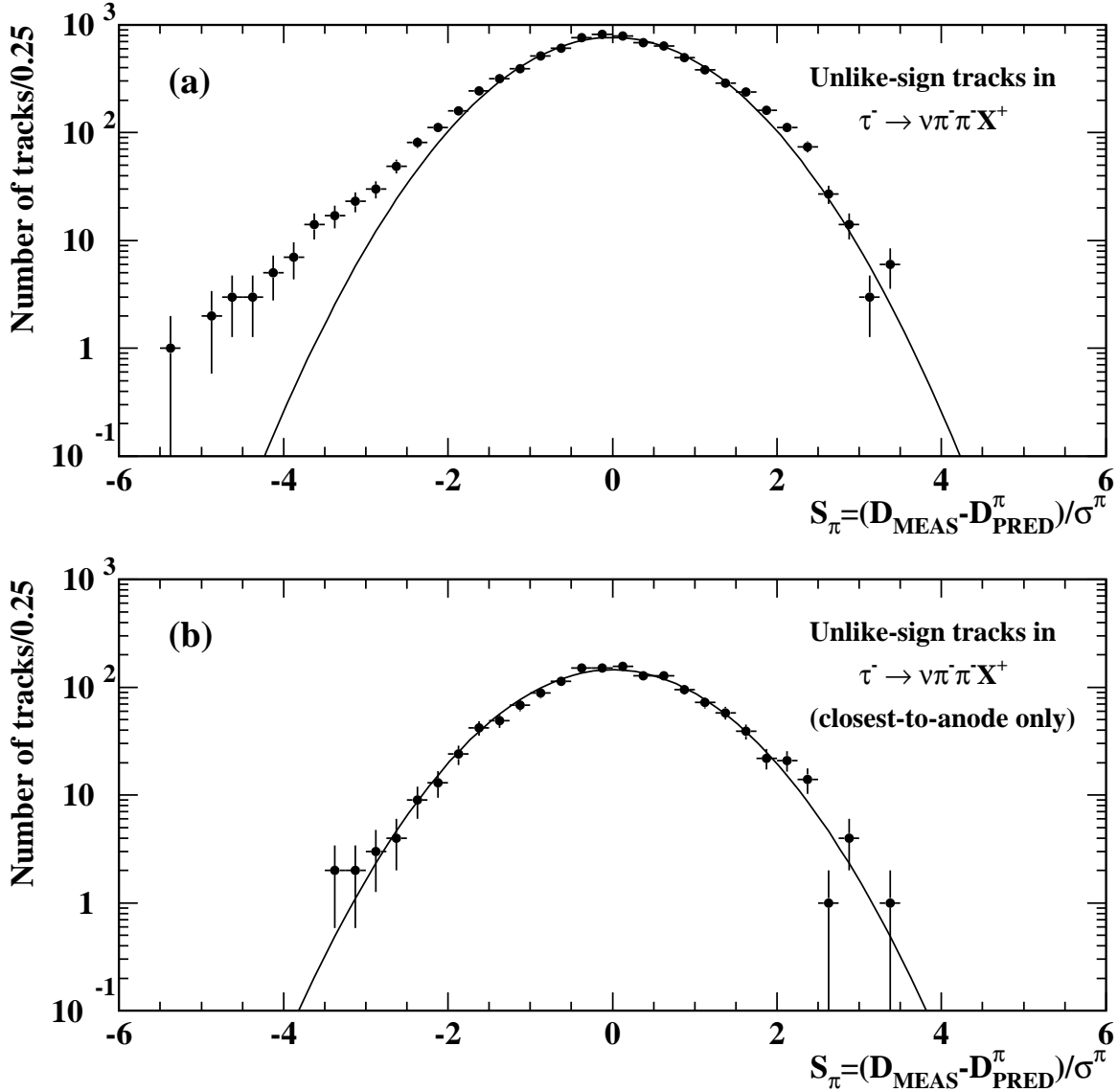} }
   \end{center}
 \vspace*{0.5cm}
 \caption[foo]{
 \label{fig:prob1}
 (a) is the stretch \de\ distribution 
 under the pion hypothesis
 for the unlike-sign tracks in the data \tppxzz\ candidate
 sample (points).  This sample is expected to 
 have less than three kaons present.
 The curve is a unit width Gaussian of zero mean, whose normalisation
 is fit to the central part of the distribution.  A significant
 non-Gaussian component is evident.
 (b) is the same distribution
 for the unlike-sign tracks in the data first-track \tppxzz\ candidate
 sample.
 }
 \end{figure}

Although the application of these corrections significantly
improves the parameterisation of \de\ in a low multiplicity
environment, the energy loss
distribution contains a significant
non-Gaussian component.
This feature is most apparent
in the corrected \de\ stretch distribution 
of the unlike-sign tracks of tau decays classified
as \tppxzz\ \footnote{The unlike-sign track is the track in the three-prong decay
with charge opposite that of the initial tau.}. Stretch energy loss,
$S_i$, is defined as:
\begin{eqnarray}
S_i = (D_{\rm meas} - D^i_{\rm pred})/\sigma^i ,
\nonumber
\end{eqnarray}
where $i$ is the particle hypothesis used to calculate the predicted
\de\ and its uncertainty.
The \tppxzz\ control sample is selected 
using $S_{\rm K} > 2.0$ to 
identify three-prong tau events with two like-sign pion candidates
in the final state.
Since the \tppkzz\ final states are highly suppressed, the
unlike-sign tracks in this sample are expected to be over $99.7\%$ pure
in 
pions.
The stretch \de\ distribution under a pion hypothesis  
of this sample is seen in Figure~\ref{fig:prob1}(a).
 
The non-Gaussian component is large in the region of the
\de\ distribution expected to be populated by kaons.
Parameterisation of the non-Gaussian component would
therefore lead to very large statistical and systematic uncertainties
in the measurement of the 
\tkppzz\ and
\tkpkzz\ branching ratios.  
The tail is 
dramatically
reduced in the \de\ distribution of tracks in 
three-prong tau decays which lie closest to the anode plane,
as shown in Figure~\ref{fig:prob1}(b).      
When
tracks are spatially close together in the jet chamber cell, 
the tail of the pulse from the track closest to the anode will effectively
shift the pedestal of the pulse of the second track, leading to an
incorrect determination of deposited charge for that track,
and the observed tails in the \de\ distribution.  
The low-end
non-Gaussian tail exists in the measured energy loss of charged
particles in
\mh\ events, but is more pronounced in three-prong tau
decays due to the tighter collimation of the particles in the decay.

The simplest method to minimize this effect is to only consider
the \de\ of the track in the three-prong
tau decay closest to the anode plane.  Tracks meeting this requirement
will be referred to as the `first-track' three-prong sample.
Further, the effect of pulses following the first is reduced by only
considering first pulses separated from the second by more than 1 cm
\cite{bib:dedxref}.
 
The corrected
stretch \de\ distribution
of unlike-sign tracks in the first-track \tppx\ sample,
seen in Figure \ref{fig:prob1}(b), is consistent with
a Gaussian of zero mean and unit width.

\section{Monte Carlo Generated Event Samples} 
\leavevmode\indent
For this analysis, tau lepton decays are simulated with
the 
KORALZ $4.0$ Monte Carlo generator and the 
Tauola $2.4$ decay package \cite{bib:koralz} \cite{bib:tauola}.
To estimate biases introduced to the branching ratios
by the tau-pair pre-selection procedure, we use
a Monte Carlo sample of $600000$ tau-pair
events that
have input branching ratios determined from world averages or theoretical
expectations.
In order to 
determine the branching ratios and to study the intermediate
resonant structure of \tkpp\ and \tkpk\ decays,  
large Monte Carlo samples
of signal events are also produced.
All Monte Carlo samples are passed through a detailed
simulation of the OPAL detector \cite{bib:geant} and are
subjected to the same analysis chain as the data.
 
The \tkpp\ decay mode is generated in Tauola $2.4$ primarily through the
${\rm K}_1 (1400)$ intermediate resonance:
\begin{eqnarray}
\tau^- & \rightarrow & \nu_\tau {\rm K}_1 (1400) \rightarrow \nu_\tau {\rm K}^{\ast}(892) \pi^- \rightarrow \nu_\tau {\rm K}^-\pi^-\pi^+
\label{kpp1} \\
\tau^- & \rightarrow & \nu_\tau {\rm K}_1 (1400) \rightarrow \nu_\tau \rho(770) {\rm K}^- \rightarrow \nu_\tau {\rm K}^-\pi^-\pi^+  .
\label{kpp2}
\end{eqnarray} 
However, the resonant structure of \tkpp\ decays is likely more diverse than that
represented in (\ref{kpp1}) and (\ref{kpp2}), as the decay can also
occur through the ${\rm K}_1 (1270)$ intermediate resonance \cite{bib:suzuki}. 
The ${\rm K}_1 (1270)$ predominantly decays
to ${\rm K}^{\ast} (892) \pi$, ${\rm K}^{\ast}_0 (1430) \pi$, and
$\rho(770) {\rm K}$ 
\cite{bib:PDG}. To assess the sensitivity of
the efficiency estimation to the assumptions made about the
\tkpp\ intermediate resonant structure, a modified
version of Tauola is used to generate the following decays:
\begin{eqnarray}
\tau^- & \rightarrow & \nu_\tau {\rm K}_1 (1270) \rightarrow \nu_\tau {\rm K}^{\ast}(892) \pi^- \rightarrow \nu_\tau {\rm K}^-\pi^-\pi^+ \nonumber \\
\tau^- & \rightarrow & \nu_\tau {\rm K}_1 (1270) \rightarrow \nu_\tau {\rm K}^{\ast}_0(1430) \pi^- \rightarrow \nu_\tau {\rm K}^-\pi^-\pi^+ \nonumber \\
\tau^- & \rightarrow & \nu_\tau {\rm K}_1 (1270) \rightarrow \nu_\tau \rho(770) {\rm K}^- \rightarrow \nu_\tau {\rm K}^-\pi^-\pi^+ \nonumber  .
\end{eqnarray}
In each case, the ${\rm K}_1 (1270)$ decays through a Breit-Wigner resonance with an
$s$-dependent width, and phase space limitations are taken into account
\cite{bib:phase}.  Other ${\rm K}_1 (1270)$ decay modes are
ignored and interference effects between the various decay chains are neglected.
 Based upon experimental measurements of the ${\rm K}_1 (1270)$ partial
widths \cite{bib:PDG}, the ${\rm K}_1 (1270)$ is assumed to decay $49\%$ of the
time to $K\rho$, $32\%$ of the time to ${\rm K}^\ast_0 (1430) \pi$, and
$19\%$ of the time to ${\rm K}^\ast (892) \pi$.
 
A sample of \tkpp\ events is also generated
through the ${\rm K}_1(1400)$ resonance, again with the
${\rm K}_1$ decaying
through a Breit-Wigner resonance with an $s$-dependent width.  We
only allow the ${\rm K}_1 (1400)$'s in this sample to decay to ${\rm K}^\ast(892) \pi$, as
experimental measurements indicate that this decay mode occurs much
more frequently than  
the $\rho(770) {\rm K}$ mode \cite{bib:k1}.

The \tkpk\ decay mode is generated in Tauola $2.4$ through a mixture of the
${\rm a}_1 (1260)$ and $\rho(1700)$ intermediate resonances:
\begin{eqnarray}
\tau^- & \rightarrow & \nu_\tau {\rm a}_1 (1260) \rightarrow \nu_\tau {\rm K}^{\ast}(892)
{\rm K}^- \rightarrow \nu_\tau {\rm K}^-\pi^-{\rm K}^+
\nonumber \\
\tau^- & \rightarrow & \nu_\tau \rho(1700) \rightarrow \nu_\tau {\rm K}^\ast(892) {\rm K}^-
\rightarrow \nu_\tau {\rm K}^-\pi^-{\rm K}^+ .
\nonumber
\end{eqnarray}
 
Since the intermediate structure of the 
$\nu_\tau {\rm K}^- \pi^- \pi^+ \pi^0$ 
and
$\nu_\tau {\rm K}^- \pi^- {\rm K}^+ \pi^0$ tau decay
final states is not well understood theoretically, and
experimental data on these decays are sparse,
the Monte Carlo samples for these channels are generated
through non-resonant production only.  
As a cross-check to 
ensure that the results of the \tkpp\ and \tkpk\ resonant
studies are independent of the modelling of the 
\tkppz\ and \tkpkz\ backgrounds, the following Monte Carlo samples
are also used:
\begin{eqnarray}
\tau^- & \rightarrow & \nu_\tau {\rm K}^- \omega \rightarrow \nu_\tau {\rm K}^-\pi^-\pi^+ \pi^0 \nonumber \\
\tau^- & \rightarrow & \nu_\tau ({\rm K}^\ast(892) {\rm K} \pi)^-  \rightarrow \nu_\tau {\rm K}^-\pi^-{\rm K}^+ \pi^0 .
\nonumber
\end{eqnarray}
 
The second-order weak \tppkzz\ decays
are modelled using Monte Carlo samples of
\tkppzz\ decays,
where the charge of the like-sign kaon and the unlike-sign pion are
reversed.  
 
The efficiency corrections in the branching ratio analysis
are estimated using \tkpp\ and \tkpk\ events generated with the
default version of Tauola~$2.4$, and \tkppz\ and \tkpkz\ events
generated through non-resonant production. 
As a cross-check, the analysis
is also performed with efficiency corrections derived from the
$\tau^-\rightarrow \nu_\tau {\rm K}_1(1270)$ and
$\tau^-\rightarrow \nu_\tau {\rm K}_1(1400)$ Monte Carlo samples.
 
\section{Branching Ratio Measurements}
\subsection{Event Selection}
\leavevmode\indent
This analysis uses full data set collected
by the OPAL detector
between the years 1990 and 1995 at \epem\ centre-of-mass energies
close to the ${\rm Z}^0$ resonance.  Only data for which
the tracking chambers
and the electromagnetic calorimeter were fully operational are retained.
The topology of \eett\ events is characterised by a pair of back-to-back,
narrow jets with low particle multiplicity.  Jets are defined in this
analysis by grouping tracks and electromagnetic clusters into
cones with an opening half-angle of $35^\circ$, where each cone is assumed
to contain the decay products of one of the tau leptons.
Tau-pair candidates are selected by requiring two low-multiplicity
jets with an average polar angle of $|\cos{\theta_{\rm ave}}| < 0.90$.
Background events from other lepton pairs, \mh\ decays, 
and two-photon events
are reduced with cuts on the event topology and total visible
energy.  These selections 
produce a sample of 147926 tau-pair candidates,
with background  
$f^{{\rm non-}\tau} = 2.73 \pm 0.27 \%$, 
estimated by data control samples and Monte Carlo background
samples. 
The tau-pair selection procedure is the same as that which
is described in detail in reference \cite{bib:john}, except for
the average polar angle selection, which is extended in this
analysis to include tracks in the endcap region of the OPAL
detector.     

The sum of the charges of the particles in each 
pre-selected tau-pair candidate event is 
required to be equal to zero.  
Three-prong tau decay candidates
are selected from the jets in this sample containing
three
well-reconstructed charged tracks.  The sum of the charges of the
three tracks is required to be $\pm 1$, and decays with
${\rm K}^0_{\rm S}$ candidates are excluded 
by rejecting tau decay candidates with one well-reconstructed
neutral vertex with $\pi^+\pi^-$ invariant mass
between $0.4$ and $0.6$ GeV \cite{bib:monica}. 
The charge of the initial tau
is taken to be the sum of the charges of the tracks in the cone
associated with the tau lepton decay.
This selection yields 38995 three-prong tau decay candidates.
 
\subsubsection{Inclusive Candidate Samples (no $\pi^0$ Identification)}
The `first-track' sample is formed by selecting the
track in each three-prong tau decay candidate that is closest
to the anode plane.  This track is also required to lie within
$|\cos{\theta}| < 0.90$, and
have at least 40 jet chamber hits used in the measurement
of the \de.  All tracks in the first-track sample
are also required to have reconstructed momentum between 3 and 90~ GeV/c.
Tracks in the barrel region of the OPAL detector 
($|\cos{\theta}| < 0.72$)
are required to have
at least 3 hits in the $z-$chambers.
Tracks which cross another track within the volume of the OPAL jet chamber
are rejected from this sample.
The `like-sign' and `unlike-sign'
track samples are formed by selecting tracks
in the first-track sample that have the same and opposite
charge as that of the
initial tau, respectively. The number of tracks in these
two samples are 9296 and 4722, respectively. 

In first-order weak decays of the tau
only the  
\tkpkzz\ decay modes contribute kaons to the unlike-sign
sample, whereas both the \tkpkzz\ and \tkppzz\ decay modes
contribute kaons to the like-sign sample.  
Thus, after corrections for efficiencies and subtraction of background 
kaons,   
the numbers of kaons in these two samples are
used to deduce the branching ratios to the
\tkppzz\ and \tkpkzz\ final states.
 
As a consistency check, we
verify that the \tppkzz\ final states are indeed suppressed.
To do this, we 
form a sample depleted in \tkpkzz\ decays
from the unlike-sign sample by
requiring both associated like-sign tracks to have $S_\pi > -1.0$.
The \kpke\ sample is formed from tracks in the unlike-sign sample which
do no pass this selection.
The number of tracks in these two samples is 1527 and 3195, respectively.
The numbers of kaons in the \kpks\ sample, the \kpke\ sample, and the
like-sign sample 
are
used to simultaneously determine the tau branching
fractions to the \kppzz, \kpkzz, and
\ppkzz\ final states.

\subsubsection{$\pi^0$ Identification}
\begin{figure}[ht]
   \begin{center}
     \mbox{ \epsfxsize=15cm
            \epsffile[40 155 524 647]{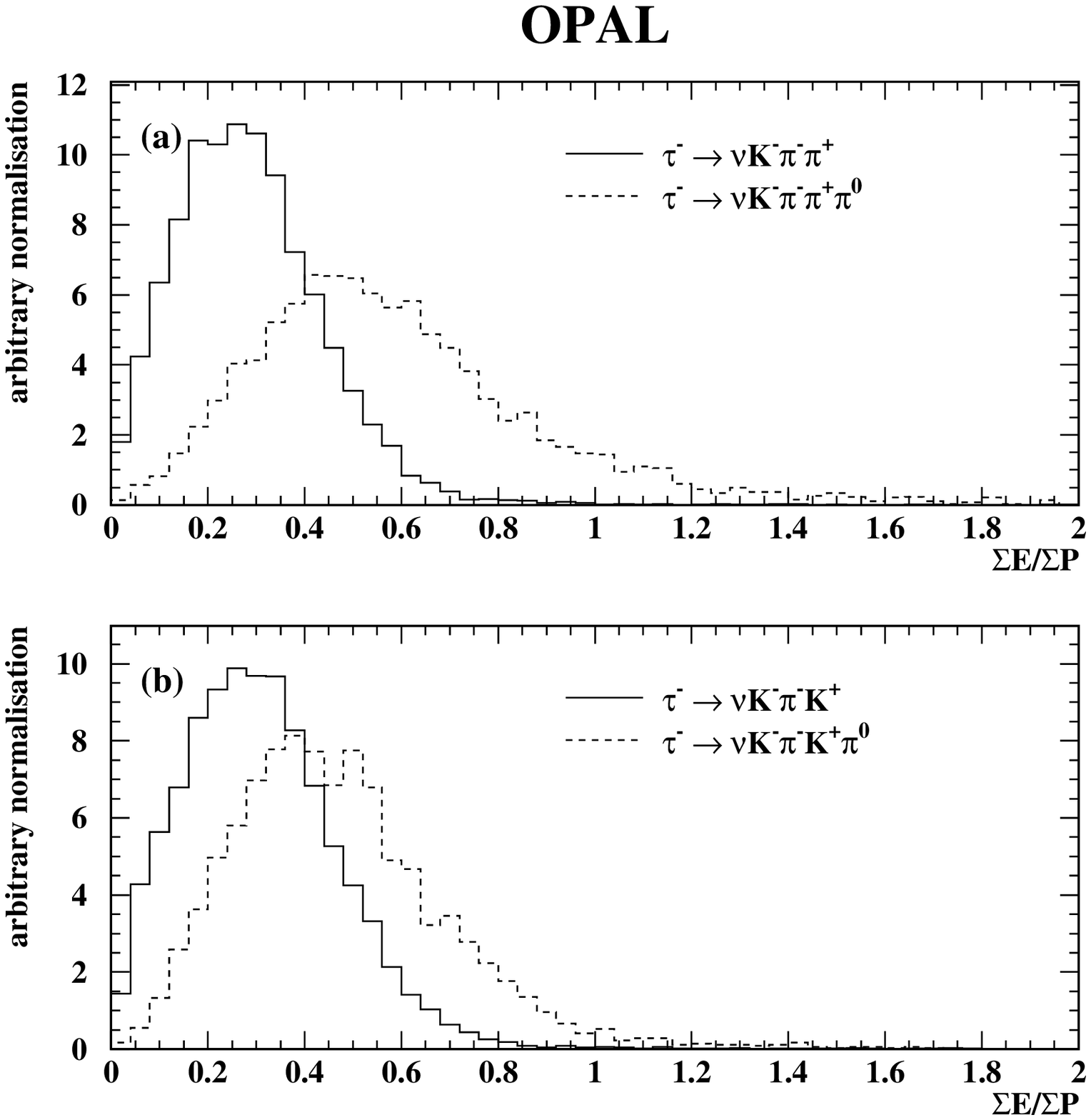} }
   \end{center}
 \vspace*{0.5cm}
 \caption[foo]{
 \label{fig:etot}
   (a) shows the 
$\sum\! E/\sum\! P$ distributions for Monte Carlo generated
\tkpp\ and \tkppz.
   (b) shows the 
same distributions for Monte Carlo generated
\tkpk\ and \tkpkz.
 }
 \end{figure}

To determine the branching ratios
for decays with and without $\pi^0$'s,
we divide all the inclusive 
samples into separate
$\pi^0$-enhanced and $\pi^0$-depleted samples using 
$\sum\! E/\sum\! P>0.5$ and $\sum\! E/\sum\! P\le0.5$, respectively,
where $\sum\! E/\sum\! P$ is the sum of the energy deposited
in the electromagnetic calorimeter, divided by the scalar
sum of the momentum of the charged particles in the tau decay cone.
The $\sum\! E/\sum\! P$ distributions for the Monte Carlo 
\tkppzz\ and \tkpkzz\ samples are
shown in Figure \ref{fig:etot}.
The number of tracks in the $\pi^0$-depleted samples with the like-sign,
unlike-sign, and \kpks\ selections
are 5984, and 3188, and 1110, respectively.
 
\subsection{Estimation of the Number of Kaons in the Samples}
\leavevmode\indent
To determine the number of kaons in the data candidate samples
described in Sections $4.1.1$ and $4.1.2$,               
we maximise a likelihood function based on \de:
\begin{eqnarray}
{\cal{L}} = \prod_{j = 1,N} \; 
            \sum_{i = {\rm e},\pi,{\rm K}}
            f_i \; W^{ij} ,
\label{eqn:like}
\end{eqnarray}
where $W^{ij}$ is the \de\ weight of charged particle $j$ under particle
hypothesis $i$,
\begin{eqnarray}
W^{ij} =    {1\over{\sqrt{2\pi} s^\prime(\beta_j) s_{\rm res} \sigma_{ij}}}
            \;
            \exp{
            \left[ - {{1}\over{2}}   
            \left( 
                { {D_{\rm meas}^j - f(\phi_j) - s^\prime(\beta_j) D_{\rm pred}^{ij}} 
                \over 
                {s^\prime(\beta_j) s_{\rm res} \sigma_{ij}} } 
            \right)^2
            \right]
            } ,
\label{eqn:like2}
\end{eqnarray}
and where 
\begin{description}
\item[{\rm $f_i$ }] is the fraction of particle type $i$ in the sample,
where $i$ is either pion, kaon, or electron.  The values of $f_i$ are
constrained to be non-negative, and the sum is constrained to be 1.
\item[{\rm $N$ }] is the total number of particles
in the sample, $N = (N_{\rm e} + N_\pi + N_{\rm K} )$.
\item[{\rm $D_{\rm meas}^j$ }] is the measured \de\ of the $j^{\rm th}$ 
charged particle.
\item[{\rm $D_{\rm pred}^{ij}$ }] is the predicted \de\
for charged particle $j$, calculated with the
OPAL parameterisation under particle hypothesis $i$ ,
as derived from the measured \de\ of charged particles in \mh\ events.
\item[{\rm $\sigma_{ij}$ }] is the uncertainty on $D_{\rm pred}^{ij}$, 
calculated
using the OPAL parameterisation,
as derived from \mh\ events.
\item[{\rm \sres }] is the multiplicative correction to $\sigma_{ij}$
for all particle hypotheses $i$, and is
allowed to vary without constraint in each fit. 
Separate \sres\ factors are used
for charged particles in the barrel and endcap regions of the OPAL detector.
The central values
of these parameters are about $0.84$ and $0.73$, respectively.
\item[{\rm \sbetap }] is the $\beta$ dependent multiplicative correction
to \dpred, and is equal to \sbeta$+\alpha$,  where \sbeta\
is determined from the one-prong control samples, 
and $\alpha$ is a correction allowed to vary in each fit
to compensate for possible slight differences between the
\de\ of the one-prong and three-prong decay environments.
Separate \sbetap\ factors are
needed for charged particles from three-prong tau decays
in the barrel and endcap regions
of the OPAL detector.
These extra constant terms yield a correction to the
\de\ of approximately $1\%$.  
\item[{\rm \fphi }] is the $\phi$ dependent correction to the
measured \de, as obtained from the one-prong control samples.
\end{description}
 
Efforts are made to obtain a \de\
parameterisation for the tau decay environment
that is optimal for many particle species
over a wide range of momenta.  However, it is possible that
the \de\ corrections described
in Section~$2.1$ may be somewhat
more (or less) optimal for pions than they are for
kaons in the momentum range of interest.
Thus, to correct for any possible
species-dependent quality differences
in the parameterisation of \de,
an
extra factor, $C_{\rm K}$, is allowed to multiply
the kaon predicted energy loss, $D_{\rm pred}^{\rm K}$, in the likelihood
function found in Equation~\ref{eqn:like}. This factor is determined
in a likelihood fit to
the measured \de\ of one-prong tau decays to hadron final states
to be
$C_{\rm K} = 0.9968 \pm 0.0014$.
 
Independent likelihood fits are performed for different
ranges of momentum
(13 bins of variable size between 3 and 90 GeV/c).
A test of the fit with large Monte Carlo 
samples verifies that the resulting estimates for the
kaon fraction have biases within the range $-0.5\sigma_{f_{\rm K}}$
to $0.3\sigma_{f_{\rm K}}$
at the $95\%$ confidence level, where
$\sigma_{f_{\rm K}}$ is the typical statistical
uncertainty on the kaon fraction
returned by the fit.  Thus, biases resulting from
the fit procedure are neglected.

\begin{figure}[ht]
   \begin{center}
     \mbox{ \epsfxsize=15cm
            \epsffile[40 155 524 647]{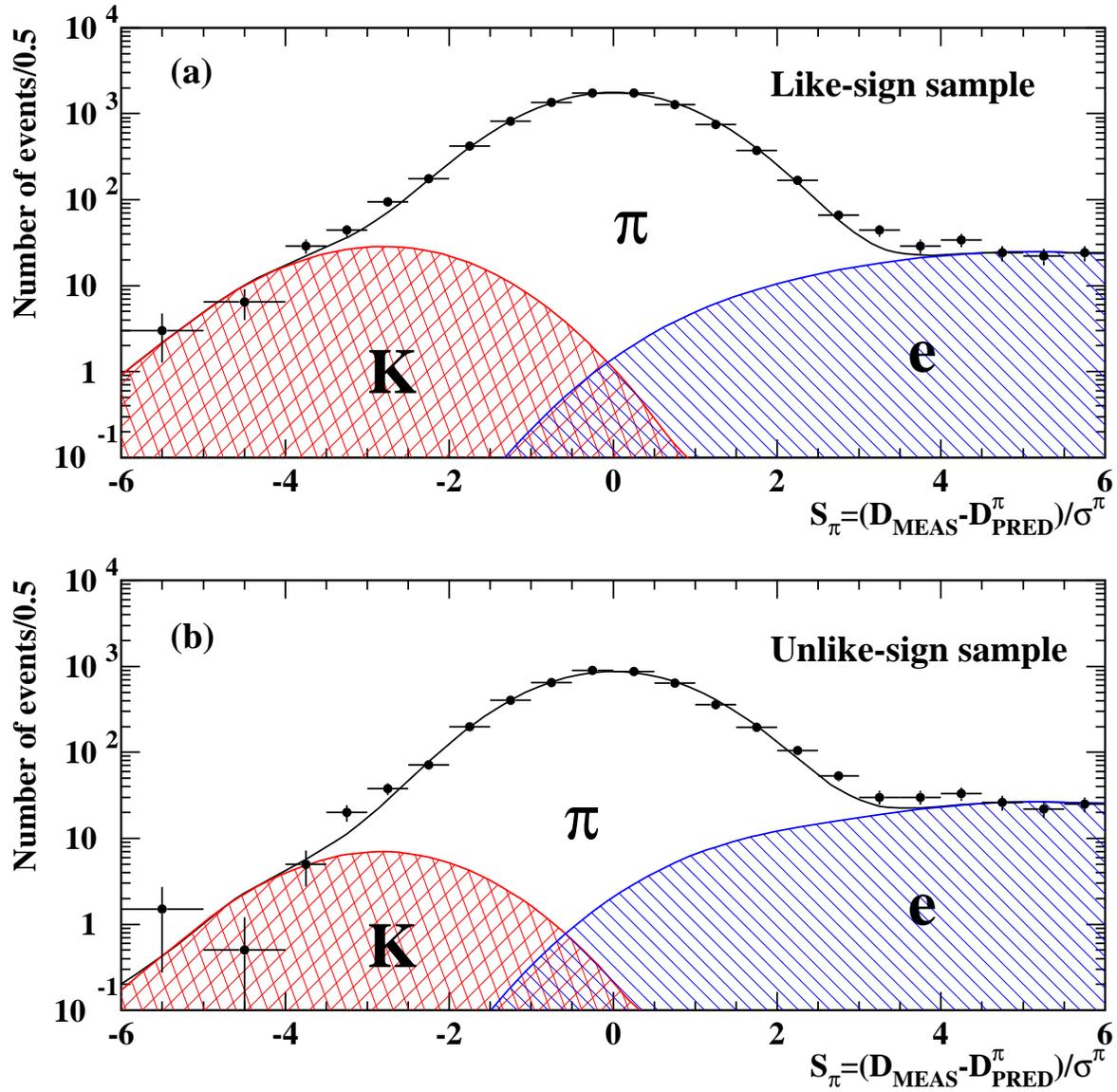} }
   \end{center}
 \vspace*{0.5cm}
 \caption[foo]{
 \label{fig:fit1}
  (a) shows the
  stretch \de\ distribution of tracks in the data inclusive
  like-sign sample (points).
  The overlaid curves are the predicted distributions for the  kaons, pions
  and electrons in the sample, assuming a Gaussian resolution
  function for the \de\ measurements.
  The normalisation of the curves comes
  from the results of the likelihood fit.
  (b) shows the same distribution for tracks in 
  the data
  inclusive unlike-sign sample.
 }
 \end{figure}

The number of kaons found within each of the samples is given in
Table \ref{tab:sum1}, and
the stretch \de\ distributions of tracks in all momentum
bins of the inclusive
data like-sign and unlike-sign samples are shown
in Figure \ref{fig:fit1}.  The
normalisation of the predicted distributions of the kaons, pions, and
electrons in each sample is obtained from the results of the likelihood
fits.  The $\chi^2$ per degree of freedom between the 
first 11 bins 
of the data and
predicted distributions, 
which is the portion of the \de\ distribution populated by kaons,
is $15.9/10$ and $10.6/7$ for the like-sign
and unlike-sign samples, respectively.

\subsection{${\rm d}E/{\rm d}x$ Systematic Studies}
\leavevmode\indent
A significant source of systematic uncertainty in this measurement
is the uncertainty in the parameterisation 
of the predicted energy loss. 
To assess this systematic,
we determine the sensitivity of the likelihood estimates of the
number of kaons within
each of the data samples 
to the uncertainties in the 
\de\ correction factors obtained from the one-prong samples.
To achieve this, we modify
the likelihood function from equation~\ref{eqn:like}
such that:
\begin{eqnarray}
 {\cal{L}}^\prime = \exp{(-{\textstyle {1\over2}}\, {\bf n^T V^{-1} n})} \; {\cal{L}} ,
\nonumber
\end{eqnarray}
where 
\begin{description}
\item[${\bf n}$] is a vector containing the difference from the
central values of the seven
\de\ correction factors that describe the function $f(\phi)$,
and the three correction factors that describe the function
$s(\beta)$, where the central values are
as derived from the fits to the one-prong control samples.
\item[${\bf V}$] is the covariance matrix 
for the \de\ correction factors.
\end{description}

In the
first iteration, the correction
factors are allowed to vary in the fit, and the returned values are
found to be consistent with the input values.
In the second iteration, the likelihood fit is repeated, this time keeping the
\de\ correction
factors fixed to the values from the first iteration.  The
systematic uncertainty in $f_{\rm K}$ due to the
parameterisation of \de\ is then obtained from the
square root of the 
quadrature difference of the fit uncertainties in $f_{\rm K}$
from the two iterations, and is shown in
Table~\ref{tab:sum1}.
 
\subsection{Efficiency Correction}
\leavevmode\indent
The efficiencies for kaons from signal events 
in the pre-selected tau-pair sample to enter
the candidate samples are estimated
using signal events generated with the KORALZ $4.0$ Monte Carlo generator and the 
Tauola $2.4$ decay package, as described in Section 3.  

Table~\ref{tab:sum1} shows the average efficiency estimates for the
various signal channels
in the pre-selected tau decay sample
that contribute kaons to the data samples described in Sections
$4.1.1$ and $4.1.2$.
All efficiencies are
corrected for biases introduced by the tau-pair pre-selection using
a Monte Carlo tau-pair sample, and
the efficiency uncertainties include the systematic 
uncertainty arising from
this correction.  The preselection biases are 
$0.935\pm0.012$, $0.925\pm0.013$, $0.908\pm0.012$,
and $0.898\pm0.013$ for \tkpp,\tkpk,\tkppz, and \tkpkz\ decays,
respectively.

\begin{figure}[p]
   \begin{center}
     \mbox{ \epsfxsize=15cm
            \epsffile[20 161 523 647]{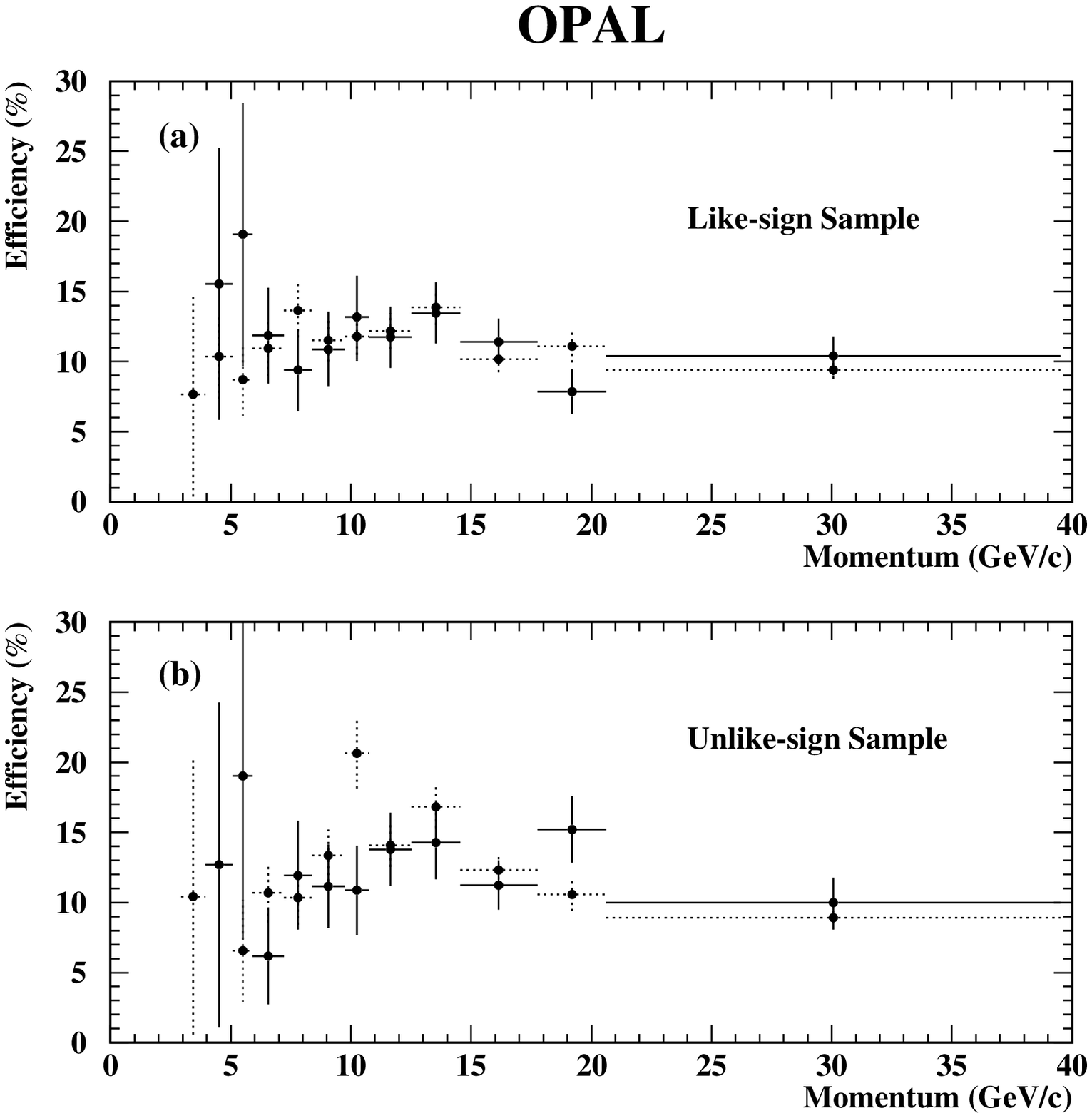} }
   \end{center}
 \vspace*{0.5cm}
 \caption[foo]{
 \label{fig:eff}
 (a) shows the 
 efficiency as a function of kaon momentum for \tkpp\  
 in the 
 pre-selected tau-pair sample to contribute kaons
 to the like-sign sample;                                        
 the solid points represent the
 default Tauola~$2.4$ generation of \tkpp\, and the
 dotted points are calculated using
  \tkpp\ generated through phase space only.
 (b) shows the 
 efficiency versus kaon momentum for \tkpk\ 
 in the
 pre-selected tau-pair sample to contribute kaons
 to the unlike-sign sample;
 the solid points represent the
 default Tauola~$2.4$ generation of \tkpk\, and the
 dotted points are calculated using
  \tkpk\ generated through phase space only.  The branching
 ratio central values are calculated using \tkpp\ and
 \tkpk\ efficiencies estimated with the default version
 of Tauola~$2.4$.
 }
 \end{figure}

Figure \ref{fig:eff} shows the efficiency versus
momentum for \tkpp\ and
\tkpk\ in the pre-selected tau-pair sample to contribute kaons to
the inclusive like-sign and unlike-sign samples, respectively.
The branching ratio analysis is performed using
an efficiency correction that is binned in momentum.
 
The variation of the branching ratios due
to alternative intermediate resonant structure scenarios for the \tkpp\
final state is assessed using efficiency estimates from   
signal events generated by the modified version of
Tauola~$2.4$ through the K$_1$(1270) and K$_1$(1400) intermediate
states as described in Section 3.  The mixture of K$_1$(1270)
and K$_1$(1400) is taken from the results of the analysis
of the intermediate resonant structure of \tkpp\ candidates, 
as described in Section 5, 
${R = {\rm Br}(\tau^-\rightarrow \,\nu_\tau\, {\rm K}_1(1270))/
{\rm Br}(\tau^-\rightarrow \,\nu_\tau\, ({\rm K}_1(1400)\;\mbox{\rm or}\; {\rm K}_1(1270)))
      =   0.71 \pm 0.19}$.
The branching ratios obtained within this range
are found
to be in agreement with the 
central values to within $1.5\sigma$ 
of the combined Monte Carlo
statistical uncertainty.
 
As a further cross-check, the branching ratios are evaluated
using \tkpp\ and \tkpk\ efficiencies estimated using 
Monte Carlo events generated through non-resonant production 
only.  The resulting branching ratios are 
in agreement with the branching ratio central values to within
$1\sigma$ of the combined Monte Carlo statistical uncertainty.
 
\subsection{Kaon Background Correction}
\leavevmode\indent
Background kaons in the data samples described in Sections 
$4.1.1$ and $4.1.2$ are estimated by
applying the same selection criteria to
Monte Carlo samples of \mh\ decays 
and tau-pair decays.
From the number of selected events that contain kaons, the
estimated number of background kaons are derived, as summarised in
Table~\ref{tab:sum1}.  Dominant sources of background kaons include 
low-multiplicity \mh\ events, 
and
${\tau^- \rightarrow \nu_\tau {\rm K}^- \ge 0 \;{\rm neutrals}}$ decays.
 
\subsection{The Branching Ratio Calculation}
\leavevmode\indent
The branching ratios in the exclusive tau decay channels of interest
are calculated from the numbers of kaons, as estimated by the
likelihood fit, in several momentum bins of the exclusive
candidate samples, listed in Table \ref{tab:sum1}.
The number of kaons within each momentum bin $i$ of 
candidate sample $j$ is corrected for background as described
in Section~$4.5$, yielding
\begin{eqnarray}
 R_{ij} = {{(N_{\rm K}^{ij} -N_{\rm bkgnd}^{ij})} 
          \over {N^{\rm pre-sel}_\tau (1-f^{{\rm non-}\tau})}} ,
\nonumber
\end{eqnarray}
where
\begin{description}
\item[{\rm $N^{\rm pre-sel}_\tau$ }] is the number of pre-selected
tau decay candidates.   There were 295852 tau decay candidates recorded
in the OPAL detector between the years
1990 and 1995.
\item[{\rm $f^{{\rm non-}\tau}$ }] is the estimated background from
non-$\tau$ sources in the pre-selected tau decay candidates
($f^{{\rm non-}\tau} = 2.73 \pm 0.27 \%$).
\item[{\rm $N_{\rm K}^{ij}$ }] is the number of kaons in momentum
 bin $i$ of
candidate sample $j$, as estimated by the likelihood fit.
 The
number of kaons summed over all momentum bins of each sample is
shown in Table~\ref{tab:sum1}.
\item[{\rm $N_{\rm bkgnd}^{ij}$ }] is the estimated number of background
kaons in momentum
 bin $i$ of candidate sample $j$,
as estimated from Monte Carlo tau decay and
\mh\ decay samples.
The
number of background kaons summed over all momentum bins of each sample is
shown in Table \ref{tab:sum1}.
\end{description}
 
A set of linear equations that relate the $R_{ij}$ to the tau branching
ratios are solved simultaneously to determine the branching
ratios.  The set of equations includes efficiency corrections
as a function of momentum
for each decay channel, as determined from Monte Carlo generated
events, as described in Section~$4.4$.  The efficiency correction
does not assume that the momentum spectrum of the signal kaons
in each sample follows those of kaons from the Monte Carlo.
For background corrections, however, the momentum spectra of
Monte Carlo generated events are used.
 
  
In the first step of the branching ratio calculation, the
\tppkzz\ branching ratios are not assumed to be zero.  
The
branching ratio calculation uses the 
division of the unlike-sign samples into the \kpkb-enhanced
and \kpkb-depleted samples in conjunction with the like-sign
samples, for a total of six exclusive
samples, as listed in Table~\ref{tab:sum1}.
The results of the calculation are:
\begin{eqnarray}
 {\rm Br}(\tau^-\rightarrow \,\nu_\tau\, {\rm K}^-\pi^-\pi^+(\pi^0)) 
     & = & +0.404\pm0.083 \;\%
\nonumber \\[1mm]
 {\rm Br}(\tau^-\rightarrow \,\nu_\tau\, {\rm K}^-\pi^-{\rm K}^+ (\pi^0))  
     & = & +0.066\pm0.079 \;\%
\nonumber \\[1mm]
 {\rm Br}(\tau^-\rightarrow \,\nu_\tau\, \pi^-\pi^-{\rm K}^+ (\pi^0))  
     & = & +0.077\pm0.070 \;\%
\nonumber \\[2mm]
 {\rm Br}(\tau^-\rightarrow \,\nu_\tau\, {\rm K}^-\pi^-\pi^+ \pi^0) 
     & = & +0.064\pm0.086 \;\%
\nonumber \\[1mm]
 {\rm Br}(\tau^-\rightarrow \,\nu_\tau\, {\rm K}^-\pi^-{\rm K}^+ \pi^0)  
     & = & -0.110\pm0.128 \;\%
\nonumber \\[1mm]
 {\rm Br}(\tau^-\rightarrow \,\nu_\tau\, \pi^-\pi^-{\rm K}^+ \pi^0)  
     & = & +0.137\pm0.109 \;\%
\nonumber \\[2mm]
 {\rm Br}(\tau^-\rightarrow \,\nu_\tau\, {\rm K}^-\pi^-\pi^+ ) 
     & = & +0.340\pm0.086 \;\%
\nonumber \\[1mm]
 {\rm Br}(\tau^-\rightarrow \,\nu_\tau\, {\rm K}^-\pi^-{\rm K}^+ )  
     & = & +0.176\pm0.078 \;\%
\nonumber \\[1mm]
 {\rm Br}(\tau^-\rightarrow \,\nu_\tau\, \pi^-\pi^-{\rm K}^+ )  
     & = & -0.060\pm0.068 \;\% ,
\nonumber
\end{eqnarray}
where the uncertainties are statistical only, and where
the inclusive branching ratios are obtained from the sums of the
exclusive branching ratios.  Correlations between the exclusive
branching ratios are taken into account in this calculation.
The results for the branching ratios to the \ppkzz\ final
states, which can only occur as second-order weak
interactions, are consistent with zero, as expected, indicating
that there are no significant biases in the kaon identification
procedure.
 
To obtain more precise branching ratios for the first-order
weak decays, we set the \tppkzz\ branching ratios to zero
and repeat the calculation.  In this iteration, 
the calculation uses the 
unlike-sign and like-sign exclusive samples, 
for a total of four samples.
The results of this calculation are:
\begin{eqnarray*}
{\rm Br}(\tau^-\rightarrow \,\nu_\tau\, {\rm K}^-\pi^-\pi^+ (\pi^0))
     & = &+0.343\pm0.073\pm0.031 \;\%
\nonumber \\[1mm]
 {\rm Br}(\tau^-\rightarrow \,\nu_\tau\, {\rm K}^-\pi^-{\rm K}^+ (\pi^0))
     & = &+0.159\pm0.053\pm0.020 \;\%
\nonumber \\[3mm]
 {\rm Br}(\tau^-\rightarrow \,\nu_\tau\, {\rm K}^-\pi^-\pi^+ \pi^0)
     & = &-0.017\pm0.076\pm0.060 \;\%
\nonumber \\[1mm]
 {\rm Br}(\tau^-\rightarrow \,\nu_\tau\, {\rm K}^-\pi^-{\rm K}^+ \pi^0)
     & = &+0.072\pm0.085\pm0.051 \;\%
\nonumber \\[3mm]
 {\rm Br}(\tau^-\rightarrow \,\nu_\tau\, {\rm K}^-\pi^-\pi^+ )
     & = &+0.360\pm0.082\pm0.048 \;\%
\nonumber \\[1mm]
 {\rm Br}(\tau^-\rightarrow \,\nu_\tau\, {\rm K}^-\pi^-{\rm K}^+ )
     & = &+0.087\pm0.056\pm0.040 \;\% ,
\nonumber
\end{eqnarray*}
where the first uncertainties are statistical and the second are systematic.
The summary of the systematic uncertainties 
for each branching ratio is shown in Table \ref{tab:sum2}.
The correlation matrix for the exclusive branching ratios is found in
Table \ref{tab:corr}.
 
The \tkppzz\ branching ratios from the first and second calculations 
are approximately 70\% correlated, and
the \tkpkzz\ branching ratios from the first and second calculations 
are approximately 50\% correlated.
In both cases, the resulting differences in the branching ratios
between the two iterations 
are consistent to within $1.3\sigma$ of the combined statistical
uncertainties.
 
The central values of the branching ratios 
are evaluated using an $\sum\! E/\sum\! P$ selection of
$0.5$ to distinguish between states including and not including a
$\pi^0$.  The branching ratios are also evaluated using 
$\sum\! E/\sum\! P$ selections 
of $0.3$, $0.4$, $0.6$, and $0.7$.  The branching ratio 
systematic uncertainty
associated with the $\sum\! E/\sum\! P$ selection is taken as the RMS 
spread of
the five values, and is quoted in Table \ref{tab:sum2}.  

\section{Resonant Structure}
\subsection{Event Selection}
\leavevmode\indent
The selection of events used in the study of the resonant
structure begins
with the three-prong tau decay candidate sample
described in Section~4.  
Events which include an \epem\ pair from a photon
converting in the detector material are identified and
rejected on the basis of a topological conversion
finder \cite{bib:monica}.
All three tracks in each decay are required to have at least
40 jet chamber hits used in the measurement of 
\de. Tracks in the barrel region of the
OPAL detector are required to have at least 3 hits in the 
$z$-chambers.
 
Using $S_\pi<-2.0$ to identify kaons and $|S_\pi|<1.5$ to
identify pions, events are classified into the following channels:
\begin{itemize}
\item \tkpp\ candidates have one like-sign kaon, and an unlike-sign
and like-sign pion.
\item \tkpk\ candidates have one like-sign pion, and an unlike-sign
and like-sign kaon.
\item \tpppzz\ candidates have all three tracks identified as pions.
\end{itemize}
The number of kaons in the lowest momentum bin
of the candidate samples used in the branching ratio
analysis is found to be consistent with zero.  Thus, to
increase the signal to noise ratio in the samples used in the
resonant studies, 
all kaon candidates are required to have momentum
of at least 5 GeV/c. Pion candidates are required to have  
momentum of at least $0.75$ GeV/c.  The above selections produce
950 \tkpp\ and 79 \tkpk\ candidates.                                    

The non-Gaussian tails in the data \de\ distribution are
not as problematic in this portion of the analysis as they are
in the branching ratio analysis and thus no attempt is made
to remove them.  
The differing shapes of the signal and the \tpppzz\ invariant
mass distributions yield some separation power between signal
and background in the data samples, and the \de\ selections 
used
to identify pions and kaons in this study 
are
designed only to provide three-prong tau
decay samples enhanced with \tkpp\ or \tkpk\ decays.
Other than these initial selections, \de\ is not
used to distinguish between signal and background.
 
\subsection{\tkpp}
The background in the \tkpp\ sample consists
primarily of \tpppzz\ events, along
with contributions from \tkppz,
other tau decays, and \mh\ events.
 
To analyse the resonant structure of these decays,
we examine
the \mkpp, \mkp, and \mpp\ invariant mass distributions
of the candidates.
To estimate
the shape of the dominant three-pion 
background in these distributions, we use
the data \tpppzz\ candidate sample, with the 5 GeV/c momentum
selection
placed on the particle which corresponds to the like-sign kaon
candidate.   This selection produces a 
sample of about 16000 events.
The shape of the \tkppz\ background in the
distributions is parameterised with Monte Carlo \tkppz\
events generated through non-resonant production.
Other background fractions are expected
to be on the order of a percent or less and are 
neglected.            
 
The description of the shape of the signal portion of
these invariant mass
distributions depends upon assumptions made
about the intermediate resonant structure of \tkpp\ decays.
We consider four different scenarios:
\begin{enumerate}
\item The \tkpp\ decays occur through a mixture
      of K$_1$(1270) and K$_1$(1400).  For this study
      we assume that the K$_1$(1270) has a
      width of $90$ MeV, and the K$_1$(1400) 
      has a width of $174$ MeV, the current world
      average widths of these resonances \cite{bib:PDG}.
\item The \tkpp\ decays occur through
      the K$_1$(1270) and K$_1$(1400) resonances, 
      and
      that the widths of these resonances are both 300 MeV,
      as suggested in reference
      \cite{bib:theory1}.
\item The \tkpp\ decays occur through the
      ${\rm K}^\ast(892)\pi^-$ and $\rho(770)K^-$ intermediate resonances.
\item The \tkpp\ decays occur through
      non-resonant production (phase space) only.
\end{enumerate}
Monte Carlo samples having 
at least 10 times the statistics of the
expected signal component of the data \tkpp\ candidates
are generated with a modified version of Tauola~$2.4$ 
under each of these resonant structure assumptions.

To estimate the background and signal fractions in
a way which accounts for correlations
between the invariant mass distributions, we divide the 
\mkpp,
\mkp, and \mpp\ invariant masses 
of the data, signal Monte Carlo sets, and background
samples 
into 7 bins each, to form a $7\!\times\!7\!\times\!7$
matrix.  
Further discrimination between  
signal and background is obtained by
dividing the scalar sum of the momenta in each three-prong
decay into 24 bins between $0$ and $48$ GeV/c.
The correlations between $\sum P$ and the
invariant masses are not significant compared
to the inter-correlations between the invariant masses.
 
Using the various signal and background distributions as templates,
a binned maximum likelihood fit is performed simultaneously
to the data $7\!\times\!7\times\!7$ matrix and the $\sum P$
distribution to 
determine
the most probable fractions of signal and \tkppz\ background in the
\tkpp\ candidates.  For the first three scenarios,
two signal fractions are allowed to float.
Bins in the matrix and $\sum P$ distribution which contain
data events but no predicted signal or background events
are neglected in the fit \footnote{Only one such bin exists in the
fit for the central values of the signal
and background fractions, and it contains only one event.}.
A test of the procedure using a Monte Carlo 
event sample of the same size and approximate composition
of the data sample reveals no significant bias
in the estimates of the signal and background
fractions from the fit.

\begin{figure}[p]
   \begin{center}
     \mbox{ \epsfxsize=15cm
            \epsffile[20 162 523 651]{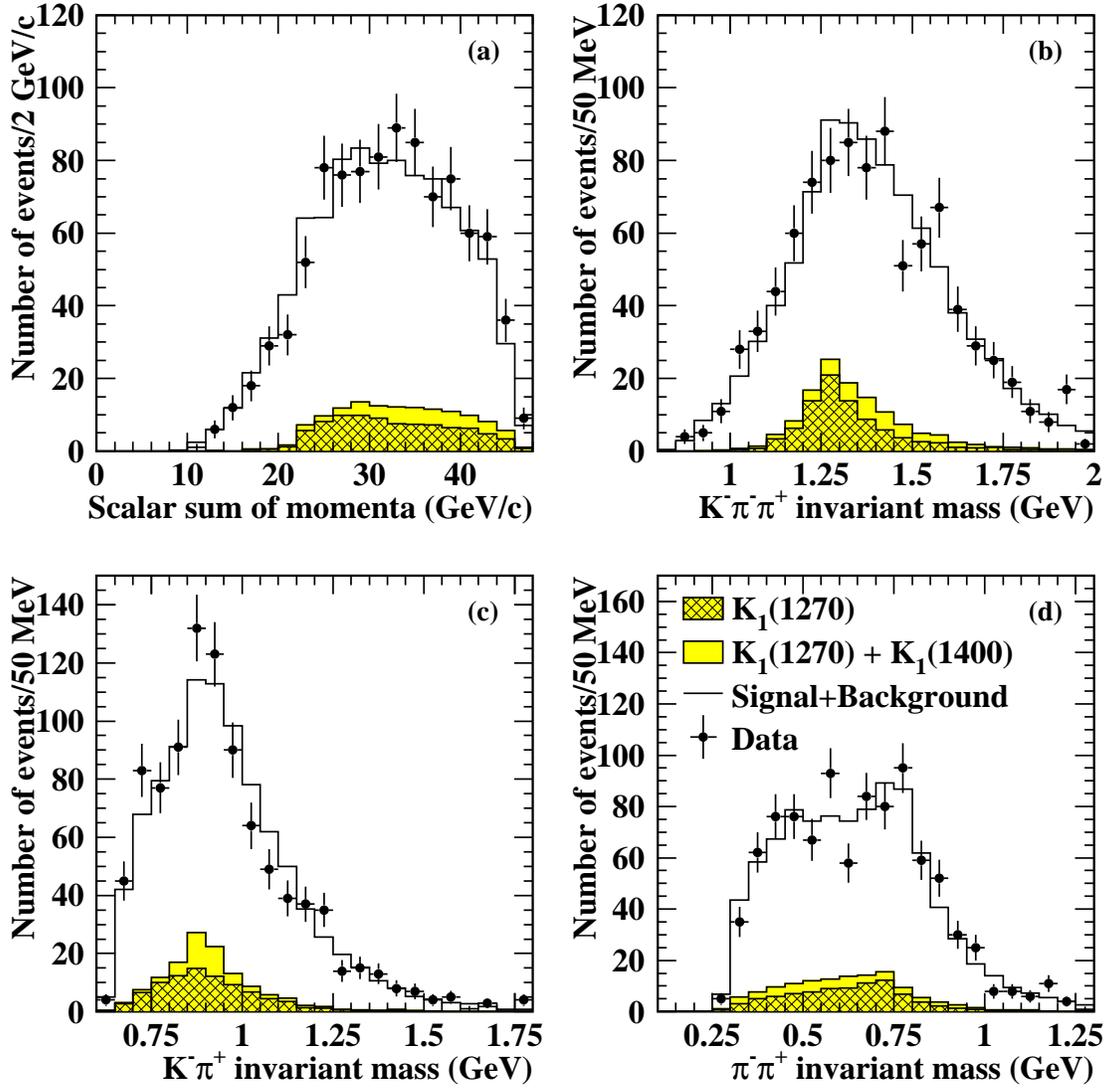} }
   \end{center}
 \vspace*{0.5cm}
 \caption[foo]{
 \label{fig:kpp2}
 (a) is the
 $\sum P$ distribution and 
 (b), (c), and (d) are the 
 invariant mass distributions 
 of the data \tkpp\ candidate
 sample (points).  
 The histogram represents the
 predicted distribution, in which 
 the normalisation of the background and signal
 components comes from the best fit
 results of the binned maximum likelihood
 fit which took the correlations between the invariant
 mass distributions into account.  
 The shape of the signal is estimated using \tkpp\ events
 generated through the ${\rm K}_1(1270)$ and the ${\rm K}_1(1400)$
 resonances, where the widths of the resonances are taken to
 be 90 and 174 MeV, respectively \cite{bib:PDG}.
 The shape of the \tpppzz\ background in the distributions is
 obtained from a data control sample.
 }
 \end{figure}

The results of the fits are shown in Table \ref{tab:kppfit}.
Table \ref{tab:kppfit} also gives the $\chi^2$ per degree of
freedom between the data and 
predicted $\sum P$, \mkpp, \mkp, and \mpp\ 
distributions, where the normalisation of the predicted 
distributions is obtained from the fit results. The
binning of these distributions is the same as that seen
in Figure \ref{fig:kpp2}, which displays the results of the
fit to the data using the assumption of the world average widths for
the two K$_1$ resonances. 
The correlation between the estimated fractions
of K$_1$(1270) and K$_1$(1400) obtained from this fit is about $-0.30$.
A fit where the signal and \tkppz\ fractions
are set to zero yields $-2\log{{\cal{L}}/{\cal{L}}_{\rm max}}=107$.
Since this fit has three fewer degrees of freedom than the first, the
difference in $-2\log{{\cal{L}}}$ should be $\chi^2$ distributed with
three degrees of freedom if the sample does indeed consist of only
three-pion background. 
This test disfavours this hypothesis 
at a confidence level of over 99\%.

Table \ref{tab:kppfit} includes estimates of the efficiencies
for \tkpp\ and \tkppz\ in the pre-selected tau-pair
sample to contribute to the \tkpp\ candidate
samples.  These efficiencies are not corrected for 
biases introduced by the \tkpp\ selection procedure.  
In order to verify that the signal fractions returned by
the fit are reasonable, we 
use these efficiencies and the fit fractions
to calculate estimates of the \tkpp\ and 
\tkppz\ branching ratios, which are also given in Table \ref{tab:kppfit}.  
Although the efficiencies have not been corrected for bias,
these estimates are
in agreement with the branching ratios obtained in Section 4
for all resonant structure assumptions,
except for \tkpp\ through phase space only. 
 
As a cross-check, Monte Carlo generated
$\tau^- \!\! \rightarrow \!\! \nu_\tau {\rm K}^- \omega$ decays are
used to estimate the shape of the \tkppz\ in the
data distributions, rather than \tkppz\ generated
through non-resonant production.  For all four 
assumptions made about the
\tkpp\ intermediate resonant structure, these fits
return signal fractions within $0.15\sigma$ of the
central values, and \tkppz\ fractions within $0.35\sigma$ of
the central values, where $\sigma$ refers to the statistical
uncertainties of the central values of the fractions.
 
\subsubsection{$\tau^-\rightarrow \nu_\tau {\rm K}_1$}
Under the assumption that resonant structure of 
\tkpp\ decays is dominated by
the ${\rm K}_1$ intermediate resonances,  we derive
\begin{eqnarray*}
 R_{\rm fit} = { {\mbox{fraction of K$_1$(1270) in \kppb\ candidates}}\over
 {\mbox{fraction of K$_1$(1400)+K$_1$(1270) in \kppb\ candidates}} } 
 & = & 0.69 \pm 0.16  ,
\nonumber
\end{eqnarray*}
where the uncertainty is statistical, and
where the fractions of ${\rm K}_1(1270)$ and ${\rm K}_1(1400)$ are taken
from the fit which assumes the widths of these resonances are
90 MeV and 174 MeV, respectively.

The signal and background
template distributions used in the binned likelihood fit
have statistical uncertainties associated with them.  To determine
the systematic uncertainties on the fit fractions arising from
these uncertainties, the fit is repeated 25 times, each time
randomly varying each bin of the template distributions by sampling
Poisson distributions with means equal to the original bin contents.
The resulting systematic uncertainties 
are included in Table \ref{tab:kppfit}, and are derived from the
RMS spread of the signal fraction estimates from the 25 fits.
 
To check for undue variation that may be produced
by the choice of binning used in the
fit, the above procedure is repeated for various binning schemes.
In all cases, the RMS variation of the results returned by the
different fits is
less than the statistical uncertainty from the original fit.
The variation of $R_{\rm fit}$ due to uncertainty in the ${\rm K}_1(1270)$
branching fractions is also studied, and found
to be negligible.
 
The value of $R_{\rm fit}$ is corrected for the efficiencies
for K$_1$(1270) and K$_1$(1400) in the pre-selected tau-pair sample
to contribute to the \tkpp\ candidate sample ($0.052\pm0.001$ 
and $0.058\pm0.002$, respectively).
This yields:
\begin{eqnarray*}
 R = {{{\rm Br}(\tau^-\rightarrow \,\nu_\tau\, {\rm K}_1(1270))}\over
 {
{\rm Br}(\tau^-\rightarrow \,\nu_\tau\, {\rm K}_1(1400))
+
{\rm Br}(\tau^-\rightarrow \,\nu_\tau\, {\rm K}_1(1270))}
}
     & = &  0.71 \pm 0.16 \pm 0.11 ,
\nonumber
\end{eqnarray*}
where the first uncertainty is due to the statistical uncertainty
from the fit, and the second arises from the limited statistics
of the Monte Carlo generated samples and the contribution from other
systematic effects.
 
To determine if $R$ depends on assumptions made
about the widths of the ${\rm K}_1$ resonances, the above
procedure is repeated, simulating the signal
portion of the distributions using \tkpp\ Monte Carlo events 
generated through the ${\rm K}_1(1270)$ and ${\rm K}_1(1400)$ resonances,
both with width 300 MeV.  This procedure yields:
\begin{eqnarray*}
 R  
     & = &  0.68 \pm 0.13 \pm 0.11 ,
\nonumber
\end{eqnarray*}
which is in agreement with the central value of $R$ to within
$0.2\sigma$ of the combined Monte Carlo statistical uncertainty.

\subsection{\tkpk}
The \tkpk\ candidate sample is expected to consist
of \tpppzz\ background, along
with \tkpk\ signal, some \tkpkz\
contamination,
and some contamination
from other tau decays and \mh\ events.
 
To analyse the resonant structure of \tkpk\ decays,
we examine
the \mkpk, \mkk, and \mpk\ invariant mass distributions
of the candidates.
To estimate
the shape of the three-pion
background in these distributions, we use
the data \tpppzz\ candidate sample, with the 5 GeV/c momentum
selection
placed on the particles which correspond to the 
two kaon candidates. This selection produces a sample
of over 10000 events.
 
The shape of the \tkpkz\ background in the
distributions is parameterised with Monte Carlo \tkpkz\
events generated through phase space.
Other background fractions are expected
to be on the order of a percent or less and are
neglected.
 
To describe the shape of the signal portion of
these invariant mass
distributions,
we consider three different scenarios:
\begin{enumerate}
\item The \tkpk\ decays occur 
      through the same intermediate resonant structure
      used in the default version of Tauola~$2.4$,
      as described in Section 3.
\item The \tkpk\ decays occur through 
      ${\rm K}^\ast(892){\rm K}^-$.
\item The \tkpk\ decays occur through
      phase space only.
\end{enumerate}
Monte Carlo samples having 
at least 10 times the statistics of the
expected signal component of the data \tkpp\ candidates
are generated  
under each of these resonant structure assumptions.

To determine the composition of the
candidate sample, we divide the \mkpk,
\mkk, and \mpk\ invariant masses
of the data, signal Monte Carlo, and background
samples 
into 24 bins each.
Further discrimination between  
signal and background is obtained by
dividing the scalar sum of the momenta in each three-prong
decay into 24 bins between $0$ and $48$ GeV/c.
Since this study is qualitative in nature only, correlations
between the distributions are ignored.
 
\begin{figure}[p]
   \begin{center}
     \mbox{ \epsfxsize=15cm
            \epsffile[20 162 523 651]{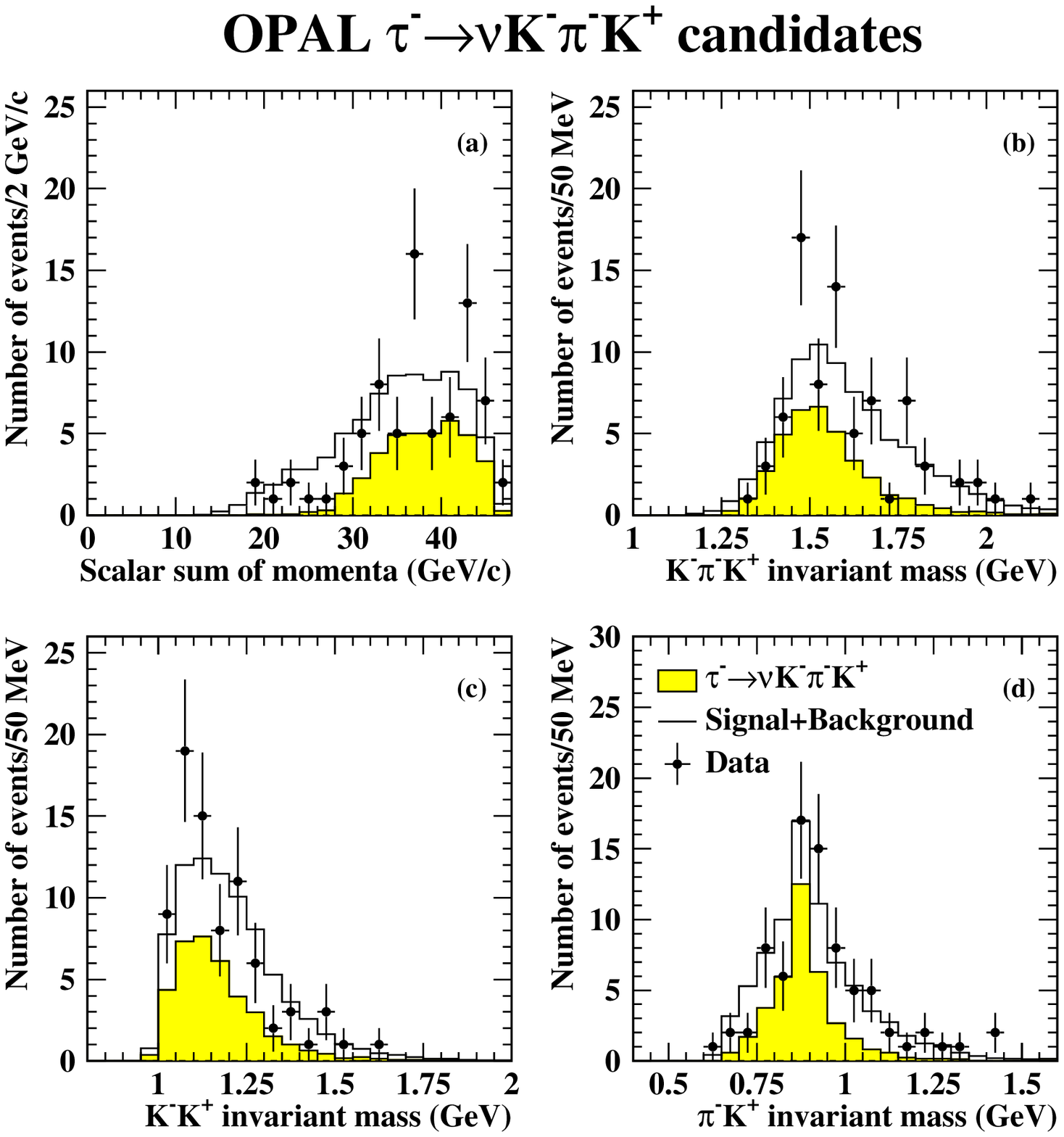} }
   \end{center}
 \vspace*{0.5cm}
 \caption[foo]{
 \label{fig:mass_kpk1}
 (a) is the
 $\sum P$ distribution and 
 (b), (c), and (d) are the 
 invariant mass distributions 
 of the data \tkpk\ candidate
 sample (points).  
 The histogram represents the
 predicted distribution, in which 
 the normalisation of the background and signal
 components comes from the results of the binned maximum likelihood
 fit to the four
 distributions. The shape of the \tkpk\ signal
 distribution is simulated in the fit by Monte Carlo events generated 
 by the default version of Tauola~$2.4$.  
 The shape of the \tpppzz\ background in the distributions is
 obtained from a data control sample.
 }
 \end{figure}

The results of the fits using the three resonant structure
assumptions are shown in Table \ref{tab:kpkfit}.
Table \ref{tab:kpkfit} also gives the $\chi^2$ per degree of
freedom between the data and
predicted $\sum P$, \mkpk, \mkk, and \mpk\
distributions, where the normalisation of the predicted
distributions is obtained from the fit results. 
The
binning of these distributions is the same as that seen
in Figure \ref{fig:mass_kpk1}, which displays the results
of the best fit to the data.
A fit where the signal and \tkpkz\ fractions
are set to zero yields $-2\log{{\cal{L}}/{\cal{L}}_{\rm max}}=51$.
Since this fit has two fewer degrees of freedom than the first, the
difference in $-2\log{{\cal{L}}}$ should be $\chi^2$ distributed with
two degrees of freedom if the sample does indeed consist of only
three-pion background. 
This test disfavours this hypothesis 
at a confidence level of over 99\%.

Table \ref{tab:kpkfit} includes estimates of the efficiencies
for \tkpk\ and \tkpkz\ in the tau-pair sample 
to contribute to the \tkpk\ candidate
samples.  These efficiencies are not corrected for 
biases introduced by the \tkpk\ selection procedure.  
In order to verify that the signal fractions returned by
the fit are reasonable, we 
use these efficiencies and the fit fractions
to calculate estimates of the \tkpk\ and 
\tkpkz\ branching ratios, which are also given in Table \ref{tab:kpkfit}.  
Although the efficiencies have not been corrected for bias,
these estimates are
in agreement with the branching ratios obtained in Section 4
for all resonant structure assumptions.
 
As a cross-check, Monte Carlo generated 
$\tau^- \!\! \rightarrow  \!\! \nu_\tau ({\rm K}^\ast(892) {\rm K} \pi)^-$ decays are
used to estimate the shape of the \tkpkz\ in the
data distributions, rather than \tkpkz\ generated
through non-resonant production.  For all three
assumptions made about the
\tkpk\ intermediate resonant structure, these fits
return signal fractions within $0.25\sigma$ of the
central values, and \tkpkz\ fractions within $0.40\sigma$ of
the central values, where $\sigma$ refers to the statistical
uncertainties of the central values of the fractions.
 
\section{Summary and Discussion}
\leavevmode\indent
From a sample of 295852 tau decays recorded in the OPAL
detector between the years 1990 and 1995,
we determine the branching ratios:
\begin{eqnarray*}
{\rm Br}(\tau^-\rightarrow \,\nu_\tau\, {\rm K}^-\pi^-\pi^+ (\pi^0))
     & = &0.343\pm0.073\pm0.031 \;\%
\nonumber \\[1mm]
 {\rm Br}(\tau^-\rightarrow \,\nu_\tau\, {\rm K}^-\pi^-{\rm K}^+ (\pi^0))
     & = &0.159\pm0.053\pm0.020 \;\%
\nonumber \\[3mm]
 {\rm Br}(\tau^-\rightarrow \,\nu_\tau\, {\rm K}^-\pi^-\pi^+ \pi^0)
     & < & 0.179 \;\% \hspace{0.5cm}\mbox{\rm (95$\%$ CL)} 
\nonumber \\[1mm]
 {\rm Br}(\tau^-\rightarrow \,\nu_\tau\, {\rm K}^-\pi^-{\rm K}^+ \pi^0)
     & < & 0.247 \;\% \hspace{0.5cm}\mbox{\rm (95$\%$ CL)}
\nonumber \\[3mm]
 {\rm Br}(\tau^-\rightarrow \,\nu_\tau\, {\rm K}^-\pi^-\pi^+ )
     & = &0.360\pm0.082\pm0.048 \;\%
\nonumber \\[1mm]
 {\rm Br}(\tau^-\rightarrow \,\nu_\tau\, {\rm K}^-\pi^-{\rm K}^+ )
     & = &0.087\pm0.056\pm0.040 \;\% ,
\nonumber
\end{eqnarray*}
where the first uncertainty is statistical and the second is systematic.

The \tkpp\ inclusive and exclusive
branching ratios are in agreement with both theory and
previous empirical measurements, as listed in Table \ref{tab:tab1a}.  
 
In a separate analysis, we explore the resonant structure
of \tkpp\ and \tkpk\ decays.
Under the assumption that 
the intermediate resonant structure of the
tau decay to the \kpp\ final state is dominated by the
K$_1$ intermediate resonances, we determine:
\begin{eqnarray*}
 R = {{{\rm Br}(\tau^-\rightarrow \,\nu_\tau\, {\rm K}_1(1270))}\over
 {
{\rm Br}(\tau^-\rightarrow \,\nu_\tau\, {\rm K}_1(1400))
+
{\rm Br}(\tau^-\rightarrow \,\nu_\tau\, {\rm K}_1(1270))}
}
     & = &  0.71 \pm 0.16 \pm 0.11 .
\nonumber
\end{eqnarray*}
 
There are two previously published results for $R$:                    
\begin{center}
\begin{tabbing}
\hspace{4.0cm} \= $R = 0.35 ^{+0.29}_{-0.21}$ \hspace{0.75cm}\= TPC/$2\gamma$ \=1994 \= \cite{bib:tpc} \\
 \> $R = 0.91 \pm 0.29$  \> ALEPH \>1999 \> \cite{bib:aleph2}, \\
\end{tabbing}
\end{center}
with an average of $0.63\pm0.21$.
The OPAL result is in agreement with this average.                  
 
It has been suggested in \cite{bib:theory2} that theoretical predictions
best match the world averages for the $\tau^-\rightarrow ({\rm K}\pi\pi)^-$
branching ratios if 
the ${\rm K}_1$ resonances are in fact wider than the current
world average widths, $\Gamma_{{\rm K}_1(1270)}=90$ MeV and 
$\Gamma_{{\rm K}_1(1400)}=174$ MeV.  From $SU(3)$ flavour
symmetry arguments, reference \cite{bib:theory2}
suggests that the actual widths of these resonances are likely to be
approximately the ${\rm a}_1(1260)$ width ($250$ MeV or
greater).  We find that the data do indeed favour 
wider ${\rm K}_1$ resonances, but that the world average widths
are also consistent with the data.

\appendix
\par
{\Large {\bf Acknowledgements}}
\par
We particularly wish to thank the SL Division for the efficient operation
of the LEP accelerator at all energies
 and for their continuing close cooperation with
our experimental group.  We thank our colleagues from CEA, DAPNIA/SPP,
CE-Saclay for their efforts over the years on the time-of-flight and trigger
systems which we continue to use.  In addition to the support staff at our own
institutions we are pleased to acknowledge the  \\
Department of Energy, USA, \\
National Science Foundation, USA, \\
Particle Physics and Astronomy Research Council, UK, \\
Natural Sciences and Engineering Research Council, Canada, \\
Israel Science Foundation, administered by the Israel
Academy of Science and Humanities, \\
Minerva Gesellschaft, \\
Benoziyo Center for High Energy Physics,\\
Japanese Ministry of Education, Science and Culture (the
Monbusho) and a grant under the Monbusho International
Science Research Program,\\
Japanese Society for the Promotion of Science (JSPS),\\
German Israeli Bi-national Science Foundation (GIF), \\
Bundesministerium f\"ur Bildung, Wissenschaft,
Forschung und Technologie, Germany, \\
National Research Council of Canada, \\
Research Corporation, USA,\\
Hungarian Foundation for Scientific Research, OTKA T-029328,
T023793 and OTKA F-023259.\\


\clearpage

\begin{table}
\begin{center}
\begin{tabular}{|l|c|ll|l|} \hline
 $\tau^-$ DECAY & Strangeness & \multicolumn{2}{c|}{EXPERIMENT} & \multicolumn{1}{c|}{THEORY}      \\
 MODE         &             & \multicolumn{2}{c|}{BR $(\%)$}  & \multicolumn{1}{c|}{BR $(\%)$}    \\ \hline
\rule[-1mm]{0cm}{6mm}
\kppzz & $-1$   & $0.275\pm 0.064$ &ALEPH98\cite{bib:aleph}      &  - \\[1mm]
       &        & $0.421\pm 0.068$ &CLEO98\cite{bib:cleo}       &  \\
       &        & $0.58^{+0.19}_{-0.18}$ &TPC/$2\gamma$94\cite{bib:tpc}  &  \\[1mm]
       &        & $0.22^{+0.17}_{-0.16}$ &DELCO85\cite{bib:delco}&  \\
       &        & $0.343\pm0.079$ &(this analysis)            & \\ \hline
\rule[-1mm]{0cm}{6mm}
\kpkzz & $0$    & $0.238\pm 0.042$ &ALEPH98\cite{bib:aleph}      &  - \\[1mm]
       &        & $0.178\pm 0.036$ &CLEO98\cite{bib:cleo}       &  \\
       &        & $0.15^{+0.09}_{-0.08}$ &TPC/$2\gamma$94\cite{bib:tpc}  &  \\  
       &        & $0.159\pm0.057$ &(this analysis)            & \\ \hline
\rule[-1mm]{0cm}{6mm}
\kpp & $-1$   & $0.214\pm 0.047$ &ALEPH98\cite{bib:aleph}   & $0.77$ \cite{bib:theory1} \\[1mm]
     &        & $0.346\pm 0.061$ &CLEO98\cite{bib:cleo}    & $0.35$ to $0.45$ \cite{bib:theory2} \\[1mm]
     &        & $0.360\pm0.095$ &(this analysis)     & $0.18$ \cite{bib:theory3} \\ 
\hline
\rule[-1mm]{0cm}{6mm}
\kpk & $0$    & $0.163\pm 0.027$ &ALEPH98\cite{bib:aleph}      & $0.20$ \cite{bib:theory1} \\[1mm]
     &        & $0.145\pm 0.031$ &CLEO98\cite{bib:cleo}       & $0.26$ \cite{bib:theory3} \\
     &        & $0.22^{+0.18}_{-0.12}$ &DELCO85\cite{bib:delco}& \\  
     &        & $0.087\pm0.069$ &(this analysis)            & \\ 
\hline
\rule[-1mm]{0cm}{6mm}
\kppz& $-1$   & $0.061\pm 0.043$ &ALEPH98\cite{bib:aleph}   & - \\ 
     &        & $0.075\pm 0.032$ &CLEO98\cite{bib:cleo}    &   \\
     &        & $<0.179$       &(this analysis)        &   \\ 
\hline
\rule[-1mm]{0cm}{6mm}
\kpkz& $0$    & $0.075\pm 0.033$ &ALEPH98\cite{bib:aleph}   & - \\ 
     &        & $0.033\pm 0.019$ &CLEO98\cite{bib:cleo}    &   \\
     &        & $<0.247$       &(this analysis)        &   \\ 
\hline
\rule[-1mm]{0cm}{6mm}
\ppkzz & $+1$   & $<0.25$ &TPC/$2\gamma$94\cite{bib:tpc}    & - \\ \hline
\kkpzz & $-2$   & $<0.09$ &TPC/$2\gamma$94\cite{bib:tpc}    & - \\ \hline
\kkkzz & $-1$   & $<0.21$ &TPC/$2\gamma$94\cite{bib:tpc}    & - \\             
       &        & $<0.019$ &ALEPH98\cite{bib:aleph} &   \\           
\hline
\end{tabular}
\end{center}
\caption[foo]{Three-prong decays of the tau lepton which include
charged kaons in the final state (charge conjugate decays
are implied). Experimental uncertainties are the combined statistical and
systematic uncertainties. Branching ratios quoted as limits are
the $95\%$ confidence limits, and 
the $(\pi^0)$ notation refers to decay modes with or without
an accompanying $\pi^0$.}
\label{tab:tab1a}
\end{table}

 
\begin{table}
 \begin{sideways}
 \begin{minipage}[b]{\textheight}
\begin{center}
\begin{tabular}{|l|c|c|c||c|c|c|} \hline
       & \multicolumn{3}{c||}{\large $\pi^0$-depleted Sample}   
       & \multicolumn{3}{c|}{\large $\pi^0$-enhanced Sample}   \\ \cline{2-7}
       & like-sign       & \multicolumn{2}{c||}{unlike-sign}    
       & like-sign       & \multicolumn{2}{c|}{unlike-sign}  \\ \cline{3-4}
                                                                \cline{6-7}
       &                 & \kpkb           & \kpkb        
       &                 & \kpkb           & \kpkb      \\
       &                 & enhanced        & depleted      
       &                 & enhanced        & depleted   \\ 
\hline
\rule[-1mm]{0cm}{6mm}
$N_{\rm K}$              & $128.4\pm{16.5}\pm{2.3}$ 
                         & $36.7\pm{9.5}\pm{1.2}$
                         & $1.4\pm{3.7}\pm{0.5}$ 
                         & $29.5\pm{9.2}\pm{1.1}$
                         & $5.9\pm{5.6}\pm{0.6}$
                         & $4.6\pm{3.7}\pm{0.3}$ \\[1mm]
$N_{\rm bkgnd}$      & $5.8\pm1.8$
                     & $0$
                     & $0$            
                     & $8.5\pm2.1$
                     & $2.5\pm1.5$
                     & $0$ \\ \hline
\ekpp\  & $0.094\pm0.003$ & $<0.001       $ & $<0.001       $          
        & $0.013\pm0.001$ & $<0.001       $ & $<0.001       $ \\
\ekpk\  & $0.078\pm0.003$ & $0.104\pm0.003$ & $0.003\pm0.001$          
        & $0.007\pm0.001$ & $0.008\pm0.001$ & $<0.001       $ \\
\eppk\  & $<0.001       $ & $0.065\pm0.006$ & $0.029\pm0.002$          
        & $<0.001       $ & $0.008\pm0.001$ & $0.005\pm0.001$ \\[1mm]
\ekppz\ & $0.044\pm0.002$ & $<0.001       $ & $<0.001       $          
        & $0.047\pm0.002$ & $<0.001       $ & $<0.001       $ \\
\ekpkz\ & $0.048\pm0.003$ & $0.053\pm0.003$ & $0.001\pm0.001$          
        & $0.030\pm0.002$ & $0.032\pm0.002$ & $<0.001       $ \\
\eppkz\ & $<0.001       $ & $0.033\pm0.002$ & $0.012\pm0.001$          
        & $<0.001       $ & $0.036\pm0.001$ & $0.011\pm0.001$ \\[1mm]
\hline
\end{tabular}
\end{center}
\caption[foo]{
The number of kaons in each data sample, $N_{\rm K}$,
where the first uncertainty is
the statistical uncertainty from the maximum likelihood fit to the
measured \de\ of the tracks in the 
sample, and the second is the systematic uncertainty 
arising from the uncertainties in the parameterisation of \de.
Also shown are the estimated backgrounds and the
average efficiencies for the various signal channels which contribute
to each sample.
}
\label{tab:sum1}
 \end{minipage}
 \end{sideways}
\end{table}


\begin{table}
 \begin{sideways}
 \begin{minipage}[b]{\textheight}
\begin{center}
\begin{tabular}{|l|r|r||r|r||r|r|} \hline
 & \multicolumn{6}{c|}{Branching Ratios ($\%$)} \\ \cline{2-7}
\multicolumn{1}{|c|}{}
                     & \bkppzz        & \bkpkzz           
                     & \bkppz         & \bkpkz            
                     & \bkpp          & \bkpk             
\\ \hline
\rule[-1mm]{0cm}{6mm}
Central value        & 
\multicolumn{1}{r}   { $0.343$}        & 
\multicolumn{1}{|r||}{ $0.159$}        & 
\multicolumn{1}{r}   { $-0.017$}        & 
\multicolumn{1}{|r||}{ $0.072$}        & 
\multicolumn{1}{r}   { $0.360$}        & 
\multicolumn{1}{|r|} { $0.087$} \\ \hline
$\sigma$ (stat)      & { ${\pm 0.073}$} 
                     & { ${\pm 0.053}$} 
                     & { ${\pm 0.076}$} 
                     & { ${\pm 0.085}$}         
                     & { ${\pm 0.082}$} 
                     & { ${\pm 0.056}$} \\
$\sigma$ (\de\ sys)  & { ${\pm 0.017}$} 
                     & { ${\pm 0.010}$} 
                     & { ${\pm 0.020}$} 
                     & { ${\pm 0.015}$}         
                     & { ${\pm 0.013}$} 
                     & { ${\pm 0.016}$} \\
$\sigma$ (MC stat)   & { ${\pm 0.023}$} 
                     & { ${\pm 0.012}$} 
                     & { ${\pm 0.029}$} 
                     & { ${\pm 0.018}$}         
                     & { ${\pm 0.029}$} 
                     & { ${\pm 0.018}$} \\
$\sigma$ ($\sum\! E/\sum\! P$ sys) & { ${\pm 0.012}$} 
                                   & { ${\pm 0.012}$} 
                                   & { ${\pm 0.049}$} 
                                   & { ${\pm 0.045}$}         
                                   & { ${\pm 0.036}$} 
                                   & { ${\pm 0.032}$} \\
\hline
\end{tabular}
\end{center}
\caption[foo]{
Summary of the branching ratio central values and sources of uncertainty.  
}
\label{tab:sum2}
 \end{minipage}
 \end{sideways}
\end{table}
\begin{table}
\begin{center}
\begin{tabular}{|l|c c c c|}
\hline
       & \bkpp\ & \bkpk\ & \bkppz\ & \bkpkz\ \\ \hline
\bkpp  & $+1.0$ & $-0.3$ & $-0.6$ & $+0.1$ \\
\bkpk  & $-0.3$ & $+1.0$ & $+0.4$ & $-0.8$ \\
\bkppz & $-0.6$ & $+0.3$ & $+1.0$ & $-0.5$ \\
\bkpkz & $+0.1$ & $-0.8$ & $-0.5$ & $+1.0$ \\
\hline
\end{tabular}
\end{center}
\caption[foo]{
Correlations between the exclusive branching ratios.
}
\label{tab:corr}
\end{table}


\begin{table}
\begin{sideways}
\begin{minipage}[b]{\textheight}
\begin{center}
\begin{tabular}{|l|l|l|l|l|} \hline
&
\multicolumn{1}{c|}{K$_1$(1270)} &
\multicolumn{1}{c|}{K$_1$(1270)} &
\multicolumn{1}{c|}{K$^\ast(892)\pi^-$}                 &
\multicolumn{1}{c|}{Phase Space}                     \\
&
\multicolumn{1}{c|}{and K$_1$(1400)} &
\multicolumn{1}{c|}{and K$_1$(1400)} &
\multicolumn{1}{c|}{and $\rho(770)K^-$}              &
                                \\ 
& \multicolumn{1}{c|}{$\Gamma_{1270} = 90$ MeV} 
& \multicolumn{1}{c|}{$\Gamma_{1270} = 300$ MeV} & & \\
& \multicolumn{1}{c|}{$\Gamma_{1400} = 174$ MeV} 
& \multicolumn{1}{c|}{$\Gamma_{1400} = 300$ MeV} & & \\
\hline

$f_{\rm signal\#1}$    & {\small $\faz=0.043\pm0.025\pm0.017$} 
                       & {\small $\faz=0.065\pm0.025\pm0.021$} 
                       & {\small $\fcz=0.100\pm0.027$} 
                       & {\small $0.029\pm0.033$}  \\
$f_{\rm signal\#2}$    & {\small $\fbz=0.096\pm0.034\pm0.028$}
                       & {\small $\fbz=0.121\pm0.039\pm0.036$}
                       & {\small $\fdz=0.040\pm0.235$} 
                       & - \\
$f_{{\rm K}^-\pi^-\pi^+\pi^0}$& {\small $0.066\pm0.028$} 
                        & {\small $0.048\pm0.036$}
                        & {\small $0.076\pm0.030$}
                        & {\small $0.082\pm0.029$} \\
$-2\log{{\cal{L}}/{\cal{L}}_{\rm max}}$     & 8 &  0&  3& 96 \\
\hline
$\chi^2$ ($\sum P$)           & 15.9/18 & 17.4/18& 17.3/18& 22.9/18\\
$\chi^2$ (\mkpp)              & 26.5/20 & 23.7/20& 26.7/20& 29.0/20\\
$\chi^2$ (\mkp)               & 22.3/17 & 19.5/17& 21.2/17& 37.0/17\\
$\chi^2$ (\mpp)               & 26.1/18 & 27.8/18& 26.3/18& 25.9/18\\
\hline
\ekpp    & $0.22$  & $0.22$ & $0.22$ & $0.21$ \\
\ekppz   & $0.17$  & $0.17$ & $0.17$ & $0.17$ \\ \hline
\bkpp ($\%$)   & $0.207\pm0.054$  & $0.271\pm0.058$ & $0.205\pm0.334$ & $0.045\pm0.052$ \\
\bkppz ($\%$)  & $0.128\pm0.054$  & $0.093\pm0.069$ & $0.147\pm0.058$ & $0.158\pm0.056$ \\
\hline
\end{tabular}
\end{center}
\caption[foo]{
Summary of the \tkpp\ signal and \tkppz\ background fractions obtained
from the binned maximum likelihood fits to the data
\tkpp\ candidate invariant mass and $\sum P$ distributions.  
The first
uncertainties are the statistical uncertainties from 
the fits.  The second uncertainties for $f_{{\rm K}_1(1400)}$ and
$f_{{\rm K}_1(1270)}$ are the systematic uncertainties
due to the limited statistics
in the Monte Carlo generated samples.
Also shown are the values of $-2\log{{\cal{L}}/{\cal{L}}_{\rm max}}$ from the
fit (relative to the best fit), and the resulting $\chi^2$ per degree
of freedom between the various data and predicted 
distributions.  
Estimates of the efficiencies for signal and \tkppz\ decays in the
tau-pair candidate sample to contribute to the \tkpp\
sample are given. 
From these efficiencies and the fraction estimates
from the fits, estimates of the branching ratios are
calculated.  
These branching ratios have not been corrected
for biases introduced by the event selection and are used as a 
cross-check of the analysis only. 
The branching ratio uncertainties are from the
statistical uncertainties of the fit fractions.
}
\label{tab:kppfit}
\end{minipage}
\end{sideways}
\end{table}

 
\begin{table}
\begin{center}
\begin{tabular}{|l|l|l|l|}\hline
&
\multicolumn{1}{c|}{Tauola~$2.4$ default} &
        & \\
&
\multicolumn{1}{c|}{mix of a$_1(1260)$} &
\multicolumn{1}{c|}{K$^\ast(892){\rm K}^-$}                 &
\multicolumn{1}{c|}{Phase Space}                     \\
&
\multicolumn{1}{c|}{and $\rho(1700)$} &
        & \\
\hline
 
$f_{{\rm K}^-\pi^-{\rm K}^+}$     & $0.473\pm0.057$
                            & $0.447\pm0.074$
                            & $0.517\pm0.086$  \\
$f_{{\rm K}^-\pi^-{\rm K}^+\pi^0}$& $0.000\pm0.304$
                             & $0.141\pm0.059$
                             & $0.109\pm0.061$ \\
$-2\log{{\cal{L}}/{\cal{L}}_{\rm max}}$          & 0&  18&  21 \\
\hline
$\chi^2$ ($\sum P$)              & 7.6/5   & 7.3/5   & 8.4/5 \\
$\chi^2$ (\mkpk)                 & 7.0/6   & 8.7/6   & 7.8/6 \\
$\chi^2$ (\mkk)                  & 5.5/6   & 10.6/6  & 8.5/6 \\
$\chi^2$ (\mpk)                  & 3.8/5   & 4.5/5   & 8.6/5 \\
\hline
\ekpk   & $0.13$  &  $0.14$ & $0.15$ \\
\ekpkz  & $0.10$  &  $0.10$ & $0.10$ \\ \hline
\bkpk ($\%$)   & $0.096\pm0.012$  &  $0.085\pm0.014$ & $0.096\pm0.016$ \\
\bkpkz ($\%$)  & $0.000\pm0.079$  &  $0.037\pm0.015$ & $0.028\pm0.016$ \\
\hline
\end{tabular}
\end{center}
\caption[foo]{
Summary of the \tkpk\ signal and \tkpkz\ background fractions obtained
from the binned maximum likelihood fits to the data
\tkpk\ candidate invariant mass and $\sum P$ distributions.
The
uncertainties are the statistical uncertainties from
the fits.  Also shown are the values of $-2\log{{\cal{L}}/{\cal{L}}_{\rm max}}$ from the
fit (relative to the best fit), and the resulting $\chi^2$ per degree
of freedom between the various data and predicted 
distributions.  
Estimates of the efficiencies for signal and \tkpkz\ decays in the
tau-pair candidate sample to contribute to the \tkpk\
sample are given. 
From these efficiencies and the fraction estimates
from the fits, estimates of the branching ratios are
calculated.  
These branching ratios have not been corrected
for biases introduced by the event selection and are used as a 
cross-check of the analysis only. 
The branching ratio uncertainties are from the
statistical uncertainties of the fit fractions.
}
\label{tab:kpkfit}
\end{table}
 
\clearpage

\newpage

\newpage
 
\newpage


 \vskip -2.0cm


\end{document}